\documentclass[twocolumn]{aastex63}

\usepackage{amsmath,amssymb,lineno}
\usepackage{newtxmath}
\usepackage{nicefrac}

\submitjournal{ApJ}

\submitjournal{ApJ}
\received{2023-12-27}
\revised{2024-05-27}
\accepted{2024-06-18}

\shorttitle{Constraints on transient sources of UHECRs}
\shortauthors{Marafico, Biteau, Condorelli, Deligny, Bregeon}

\newcommand{\dif}{\mathrm{d}}
\newcommand{\nn}{\mathbf{n}}

\begin{document}

\title{Closing the net on transient sources of ultra-high-energy cosmic rays}

\correspondingauthor{Jonathan Biteau, Antonio Condorelli, Olivier Deligny}
\email{biteau@in2p3.fr, antonio.condorelli@ijclab.in2p3.fr, olivier.deligny@ijclab.in2p3.fr}

\author[0000-0003-4604-4677]{Sullivan Marafico}
\affiliation{Université Paris-Saclay, CNRS/IN2P3, IJCLab, 91405 Orsay}

\author[0000-0002-4202-8939]{Jonathan Biteau}
\affiliation{Université Paris-Saclay, CNRS/IN2P3, IJCLab, 91405 Orsay}
\affiliation{Institut universitaire de France (IUF)}

\author[0000-0001-5681-0086]{Antonio Condorelli}
\author[0000-0001-6863-6572]{Olivier Deligny}
\affiliation{Université Paris-Saclay, CNRS/IN2P3, IJCLab, 91405 Orsay}

\author[0000-0002-6790-5328]{Johan Bregeon}
\affiliation{CNRS-IN2P3, Laboratoire de Physique Subatomique et de Cosmologie (LPSC), Grenoble}

\begin{abstract}
Arrival directions of ultra-high-energy cosmic rays (UHECRs) observed above $4\times10^{19}\,$eV provide evidence of localized excesses that are key to identifying their sources. We leverage the 3D matter distribution from optical and infrared surveys as a density model of UHECR sources, which are considered to be transient. Agreement of the sky model with UHECR data imposes constraints on both the emission rate per unit matter and the time spread induced by encountered turbulent magnetic fields. Based on radio measurements of cosmic magnetism, we identify the Local Sheet as the magnetized structure responsible for the kiloyear duration of UHECR bursts for an observer on Earth and find that the turbulence amplitude must be within $0.5-20\,$nG for a coherence length of $10\,$kpc. At the same time, the burst-rate density must be above $50\,$Gpc$^{-3}$\,yr$^{-1}$ for Local-Sheet galaxies to reproduce the UHECR excesses and below $5\,000\,$Gpc$^{-3}$\,yr$^{-1}$ ($30\,000\,$Gpc$^{-3}$\,yr$^{-1}$) for the Milky Way (Local-Group galaxies) not to outshine other galaxies. For the transient emissions of protons and nuclei to match the energy spectra of UHECRs, the kinetic energy of the outflows responsible for UHECR acceleration must be below $4\times10^{54}$\,erg and above $5\times10^{50}$\,erg ($2\times10^{49}$\,erg) if we consider the Milky Way (or not). The only stellar-sized transients that satisfy both Hillas' and our criteria are long gamma-ray bursts.
\end{abstract}


%\linenumbers

\section{Introduction} 
\label{sec:intro}

The astrophysical objects that accelerate charged particles to energies above 1\,EeV (${\equiv}\,10^{18}\,\mathrm{eV}$) have long been elusive. Yet, the mechanisms by which these nuclei are emitted and the abundance of nuclear elements in the source environments are considerably constrained by the energy spectrum and mass composition of the ultra-high-energy cosmic rays \citep[UHECRs,][]{PierreAuger:2021hun,PierreAuger:2014sui}. Data collected over the past two decades show a gradual increase with energy in the mean logarithmic mass number of UHECRs, $\langle \ln A\rangle$ \citep{PierreAuger:2014sui,Watson:2021rfb}. These observations are consistent with the fundamental expectation that electromagnetic processes accelerate charged particles of atomic mass $A$ to a maximum energy $E^{\mathrm{max}}_{Z_A}$ proportional, at least approximately, to their electric charge $Z_A$~\citep{Peters1961}.  The intensity of individual nuclear components, ranging from fully ionized He to Fe, is thus expected to drop off at the same magnetic rigidity at the sources, which is estimated to $E^{\mathrm{max}}_{Z_A}/Z_A \simeq 1-5$\,EV \citep[see, e.g.,][]{Aloisio:2013hya,Taylor:2015rla,PierreAuger:2016use}.

The low variance observed for the mass estimators, $\sigma^2(\ln A)$, imposes a nearly mono-elemental composition at all energies above 5\,EeV. The little mixture above this so-called ankle feature of the energy spectrum is reflected in a very hard spectral index for each of the nuclear species: $-1 \lesssim \gamma \lesssim 1$, for an emission spectrum that follows $\dif N/\dif E \propto E^{-\gamma}$ with a cutoff at the highest energies. Whatever the exact value of the spectral index, which depends on the systematic uncertainties affecting the various models~\citep{PierreAuger:2016use}, the hardness of the nuclear spectra differs from expectation of diffusive shock acceleration models, $\gamma \gtrsim 2$~\citep[see][for a review]{Sironi:2015oza}. Acceleration by back-and-forth bounces on approaching magnetic fields, described by the first-order Fermi mechanism, can however be reconciled with observations if the shaping of ejected spectra takes into account interactions within the sources. These interactions tend to favor the escape of high-energy nuclei relative to lower-energy nuclei~\citep{Globus:2015xga,Unger:2015laa,Biehl:2017zlw,Fang:2017zjf,Supanitsky:2018jje}. A generic prediction of such in-source-interaction scenarios is a proton spectrum softer than the nuclear spectra, due to the escape from the magnetized interaction zone of neutrons that subsequently decay in protons. The requirement of a soft proton spectrum and hard nuclear spectra escaping from the sources has been shown to be compatible with the UHECR data, in particular with the proton spectrum observed at energies below the ankle~\citep{Luce:2022awd}.

To supply the observed energy density, the production rate density of UHECRs is constrained to be ${ \mathcal{L}\simeq10^{54}\,}$erg\,Gpc$^{-3}$\,yr$^{-1}$ above $0.6\,$EeV~\citep{Luce:2022awd}. The value inferred for $\mathcal{L}$ can be used to identify the few classes of sources that radiate similarly in the electromagnetic band. Several candidates can indeed meet the energetics constraints, from sources that undergo persistent or frequent episodes of electromagnetic emission with rather low luminosity to cataclysmic high-luminosity events that occur only rarely \citep[see][for a review]{2019FrASS...6...23B}. The degeneracy between source density and luminosity is addressed in this work using evidence of localized UHECR excesses over the celestial sphere.

Sky maps of arrival directions support from an observational perspective that UHECRs are predominantly of extragalactic origin. On the one hand, the contrast of the flux distribution on the sphere above $8\,$EeV is suggestive of the pattern expected from the large-scale structure of matter within a few hundred Mpc \citep{PierreAuger:2017pzq}. On the other hand, the analysis of the arrival-direction data above 40\,EeV together with spectrum and composition data leads to the inference of a flux contribution of ${\simeq}\,10-20\,\%$ from the ${\simeq}\,40$ brightest star-forming galaxies within a radius of a hundred Mpc \citep{2024JCAP...01..022A}. This is consistent with the correlation of the flux pattern of these galaxies and UHECR arrival directions alone, as previously reported by the Pierre Auger Collaboration in~\cite{PierreAuger:2018qvk,PierreAuger:2022axr}. The distance scales sampled with increasing energy are in line with the reduction of the UHECR horizon, primarily due to interaction with the cosmic microwave background (CMB) and known as the GZK horizon \citep{1966PhRvL..16..748G,1966JETPL...4...78Z}. 

But sky maps of arrival directions contain a potentially even greater wealth of information that may break the degeneracy between frequent low-luminosity and rare high-luminosity source populations meeting the energetics requirements. For a given burst rate, the average number of sources visible at each instant depends on the effective duration of the bursts, and therefore on the magnetic wandering experienced by the UHECRs during propagation. As a result, the average number of visible galaxies hosting transient sources decreases with energy much faster than in a steady scenario. Steady and transient scenarios should therefore give rise to different sky maps. Signatures of transient scenarios have been used in the past to infer source properties and constrain propagation by supposing the detection of UHECR clustering on small angular scales \citep[see the seminal works of][]{Sigl:1997eh,Lemoine:1997ei}, to constrain the local burst-rate density assuming sources distributed uniformly in space and emitting protons only \citep[see][]{Murase:2008sa}, and to interpret potential features of the all-particle energy spectrum \citep[see][]{Miralda-Escude:1996twc}. 

The aim of this paper is to complement the existing constraints on the production rate density of UHECRs by making use of evidence of anisotropy on intermediate angular scales to bracket the luminosity and rate density of bursting sources. The common thread of this work is that each galaxy of the flux-limited sample by~\cite{Biteau:2021pru} can harbor transient episodes of UHECR production at a given rate. In Section~\ref{sec:model}, we describe the source-emission model that incorporates the transient nature of the UHECR production from each galaxy, the resulting intensities on Earth, the time delays and angular deflections undergone by particles during their propagation in the intervening magnetic fields, and the magnetic effects limiting the number of observable sources. In Section~\ref{sec:results}, we infer the best rate density to reproduce the main features of the sky map reported by the Pierre Auger and Telescope Array collaborations above ${\simeq}\,40$\,EeV~\citep{PierreAuger:2023mvf}. Finally, we show in section~\ref{sec:discussion} that for stellar-sized sources to simultaneously meet the composition, energy and density constraints as well as the Hillas criterion, the net closes in on long-duration gamma-ray bursts. 

\section{Description of the model}
\label{sec:model}

\subsection{Distribution of stellar matter}
\label{sec:galaxy_catalog}

To model the 3D distribution of galaxies hosting transient UHECR accelerators, we use the near-infrared flux-limited sample of \cite{Biteau:2021pru} that maps both star-formation rate (SFR) and stellar mass ($M_\star$) over 90\,\% of the sky. The catalog offers unprecedented cosmography of SFR and $M_\star$ within 350\,Mpc, including about 400\,000 galaxies. This number of entries is an order of magnitude larger than in the 2MASS redshift survey \citep[2MRS,][]{2012ApJS..199...26H} used in other recent UHECR studies \citep[e.g.][]{2022A&A...664A.120A, 2023arXiv230517811A}. Distances are estimated with a $50-50$ ratio of spectroscopic and photometric measurements. The catalog also provides corrections for incompleteness with increasing distance and with decreasing Galactic latitude. 

Beyond the Local Volume, i.e.\ at distances greater than 11\,Mpc, the incompleteness along the Galactic plane can not be compensated for. The galaxies above and below the so-called zone of avoidance, which covers the remaining 10\,\% of the sphere, are cloned mirror-symmetrically to fill it, following the approach of \cite{2011MNRAS.416.2840L}. The cloning procedure has a negligible impact below 50\,Mpc, but it allows us to fully reconstruct the contribution of more distant structures, including our supercluster Laniakea whose core region ($50-100$\,Mpc) is close to the zone of avoidance. The catalog of \cite{Biteau:2021pru} thus provides a coherent estimate of the density of matter in the entire 350-Mpc radius volume, with the greatest granularity to date.

\begin{figure*}[htp]
\centering
\includegraphics[width=\columnwidth]{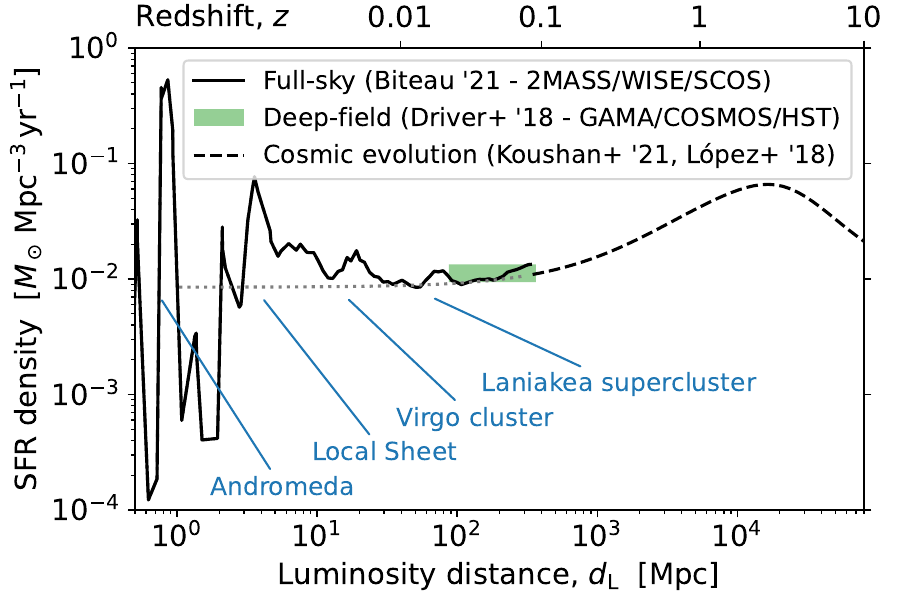}
\includegraphics[width=\columnwidth]{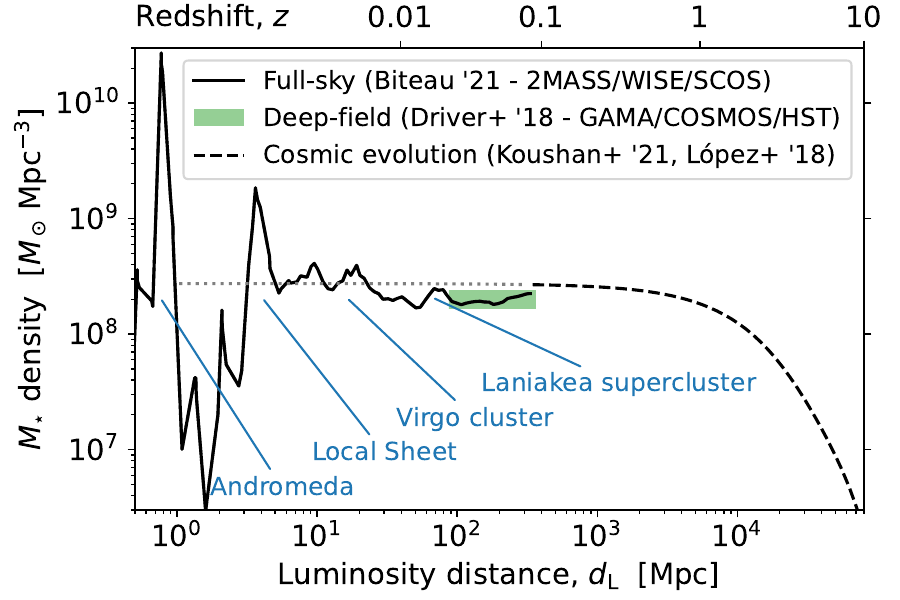}
\caption{The star-formation-rate density (\textit{left}) and stellar-mass density (\textit{right}) as a function of distance. A Chabrier initial mass function is adopted for data and models. Density profiles within a 350\,Mpc radius ($z_\mathrm{cat}^\mathrm{max} \simeq 0.08$) are calculated from the catalog of \cite{Biteau:2021pru}. The green bands illustrate the 68\,\% confidence interval of deep-field measurements at $0.02 < z < 0.08$ \citep{2018MNRAS.475.2891D}. The dashed  black lines show the cosmic evolution in best agreement with the two tracers \citep{2018A&A...615A..27L} as parametrized and normalized by \cite{2021MNRAS.503.2033K}. For comparison with measured profiles, extrapolations of cosmic evolutions at redshifts $z<0.08$ are represented by grey dotted lines.}
\label{fig:evolution}
\end{figure*}

The cosmography provided by the catalog is fully consistent with the known distribution of matter in the nearby universe. Most of the galaxies within a box of 10\,Mpc-width are in a plane defining the Local Sheet, which is tilted by  $8^\circ$ with respect to the supergalactic plane \citep{McCall:2014eha}. The $M_\star$ distribution within a box of 600\,Mpc-width matches the 3D dark-matter density fields inferred from Cosmicflows-2 \citep[see][]{Hoffman:2017ako}, which models the dark-matter density in a comparable volume using the peculiar velocities of around 8\,000 galaxies \citep[see][and reference therein for recent applications to UHECR studies on large angular scales]{2024ApJ...966...71B}. Finally, the SFR and $M_\star$ densities accounting for incompleteness corrections converge beyond 100\,Mpc towards the values measured in deep fields \citep{2018MNRAS.475.2891D}, as illustrated in Figure~\ref{fig:evolution}. 

\subsection{Source-emission model}
\label{sec:source_model}

In our model, each galaxy in the catalog is supposed to harbor UHECR sources, which burst at a given rate. We consider two tracers for the burst rate: SFR and $M_\star$. In both cases, the intensity observed on Earth receives contributions from each galaxy in the catalog within 350\,Mpc and from a continuous density of galaxies beyond 350\,Mpc. By using these two different tracers, we can show the limited influence on our results of the source evolution at high redshift ($z>0.1$).

The emission rate per unit energy  of a single galaxy $i$ is conveniently parameterized as
\begin{equation}
    \label{eqn:qAi}
    {\dot Q}_{Ai}(E) = Q^0_{A}f(E,Z_{A})S_i,
\end{equation}
with $S_i$ the ejection rate of UHECRs and $Q^0_{A}f(E,Z_{A})$ the number of particles per unit energy. The widely-used exponentially-suppressed power-law function is adopted for $f(E,Z_{A})$, as it provides a summary of the acceleration and ejection processes with only three parameters: the spectral index of protons, the spectral index of heavier nuclei and the rigidity-dependent maximum energy (see Section~\ref{sec:spectrum_compo_bestfit} for details). The emission spectrum is supposed identical at each source. Observational data indeed suggest little variation in maximum rigidity from one source to another, to preserve the mixture of elements as the mass composition gets heavier above EeV energies~\citep{Ehlert:2022jmy}. The ejection rate, on the other hand, is governed by the SFR or $M_\star$ of the galaxy
\begin{equation}
    \label{eqn:Si}
    S_i=\left\{
    \begin{array}{ll}
    k\,\frac{\mathrm{SFR}_i}{c_{N,s}(d_i, b_i)} \\
    \dot{k}\,\frac{M_{\star i}}{c_{N,m}(d_i, b_i)}
    \end{array}
    \right.,
\end{equation}
with $k$ ($\dot{k}$) expressed in $M_\odot^{-1}$ ($M_\odot^{-1}~$yr$^{-1}$) units and where $c_{N,s}$ ($c_{N,m}$) is the incompleteness correction factor. This correction factor, which depends on the luminosity distance ($d_i$) and Galactic latitude ($b_i$) of the galaxy indexed by $i$, compensates for fainter galaxies below the flux threshold of the near-infrared survey. These faint galaxies are assumed in Equation~\eqref{eqn:Si} to follow the same spatial distribution as the brighter galaxies in the catalog. In the following, it will prove useful to express ${\dot Q}_{Ai}(E)$ as
\begin{equation}
    \label{eqn:qAi-bis}
    {\dot Q}_{Ai}(E) = \frac{p_A(E)}{E}\mathcal{E}_AS_i,
\end{equation}
with $\mathcal{E}_A=Q^0_{A}\int \dif E~Ef(E,Z_A)$ the total energy emitted per source of UHECRs with mass number $A$ above a minimum rigidity $E/Z_A$. The minimum rigidity is set here at 0.6\,EV by the minimum energy of the observed proton spectrum. This way $p_A(E)=Q^0_{A}Ef(E,Z_A)/\mathcal{E}_A$ is the probability density function for the emitted energy $E$ of an UHECR with mass number $A$.

Beyond $d^\mathrm{max}_\mathrm{cat} = 350\,$Mpc, that is a redshift $z_\mathrm{cat}^\mathrm{max} \simeq 0.08$, the near-infrared survey used in \cite{Biteau:2021pru} is too shallow to resolve more than half of the matter density in individual galaxies. The ejection rate is then expressed per unit energy and per  comoving unit volume as
\begin{equation}
    \label{eqn:qA}
    {\dot q}_{A}(E,z) = q^0_{A}f(E,Z_{A})\left\{
    \begin{array}{ll}
    k\dot{\rho}_\mathrm{SFR}(z) \\
    \dot{k}\rho_{M_\star}(z)
    \end{array}
    \right..
\end{equation}
Here, $\dot{\rho}_\mathrm{SFR}(z)$ and $\rho_{M_\star}(z)$ are the star-formation-rate density (SFRD) and stellar-mass density (SMD), respectively. We adopt the cosmic evolution of the SFRD, $\dot{\rho}_\mathrm{SFR}(z)$, as provided by \cite{2018A&A...615A..27L} and normalized by \cite{2021MNRAS.503.2033K}. Of all the models tested by \cite{2021MNRAS.503.2033K}, the parametrization adopted here provides the best agreement with both SFR and $M_\star$ measurements at redshifts $z \leq z_\mathrm{cat}^\mathrm{max}$ \citep{Biteau:2021pru, 2018MNRAS.475.2891D}. The model for the cosmic evolution of the SMD, $\rho_{M_\star}(z)$, is obtained by integrating the SFRD over lookback time, the role of which is played by redshift,
\begin{equation}
    \label{eqn:rho_SMD}
    \rho_{M_\star}(z)=(1-R)\int_z^\infty \dif z'~\dot{\rho}_\mathrm{SFR}(z')\left|\frac{\dif t}{\dif z'}\right|,
\end{equation}
with $R=0.41$ the return fraction for a Chabrier initial mass function \citep{2014ARA&A..52..415M}.\footnote{The choice of initial mass function does not affect our results on the absolute density of the UHECR burst rate, expressed in units of Gpc$^{-3}$\,yr$^{-1}$.}  This fraction corresponds to the proportion of the mass of each generation of stars that is returned to the interstellar and intergalactic media. The term $|\dif t / \dif z|$ follows from the $\Lambda$CDM-cosmology parameters,
\begin{equation}
    \label{eqn:dtdz}
    \left|\frac{\dif t}{\dif z}\right|=\frac{1}{H_0(1+z)\sqrt{(1+z)^3\Omega^0_{\mathrm{m}}+\Omega^0_{\Lambda}}},
\end{equation}
with $H_0=70\,$km\,s$^{-1}$\,Mpc$^{-1}$, $\Omega^0_{\mathrm{m}}=0.3$ and $\Omega^0_{\Lambda}=0.7$.

\newpage
\subsection{Intensities on Earth}
\label{sec:intensity_model}

The all-particle energy spectrum observed on Earth, $J(E)$ in eV$^{-1}$\,km$^{-2}$\,yr$^{-1}$\,sr$^{-1}$, receives contributions from the foreground galaxies in the catalog and from an isotropic background at $z> z_\mathrm{cat}^\mathrm{max}$,
\begin{equation}
\label{eqn:J}
    J(E)=\sum_A \left[J^{\mathrm{bkg}}_A(E)+\frac{1}{4\pi}\sum_i \int\dif\nn~J^i_A(E,\nn)\right],
\end{equation}
where the outer sum is over each observed species of mass number $A$ and the inner sum is over the galaxies in the catalog. We explain below the scenario tracing the UHECR sources by the SFR of galaxies. The scenario based on stellar mass is obtained by substituting $k\rho_\mathrm{SFR}(z)$ ($k\mathrm{SFR}/c_{N,s}$) by $\dot{k}\rho_{M_{\star}}(z)$ ($\dot{k}M_{\star}/c_{N,m}$). 

Up to $z=z^\mathrm{max}_\mathrm{cat}$, each single galaxy contributes
\begin{align}
\label{eqn:model_src}
    J^i_A(E,\nn) &=  \frac{F(\nn,\nn_i;\Theta)}{4\pi d^2_i} \sum_{A'} \int \dif E' \dot Q_{A'i}(E')\frac{\dif\eta_{AA'}(E,E',z)}{\dif E} \nonumber \\
    &= \frac{\mathrm{SFR}_i}{4\pi d^2_i c_{N,s}(d_i, b_i)} F(\nn,\nn_i;\Theta) \sum_{A'} k\mathcal{E}_{A'} \nonumber \\ 
    &\quad \times \int \frac{\dif E'}{E'} p_{A'}(E')\frac{\dif\eta_{AA'}(E,E',z)}{\dif E}
\end{align}
to the observed spectrum for a given species. The factor $\mathcal{E}_{A'}$ is defined here as the energy radiated isotropically in each emitted nuclear species $A'$. The energy losses and spallation processes are described by $\eta_{AA'}(E,E',z)$, which is the fraction of particles detected on Earth with energy $E$ and mass number $A$ from parent particles emitted by the sources with energies $E' \geq E$ and mass numbers $A' \geq A$. In practice, $\eta_{AA'}(E,E',z)$ is calculated by Monte Carlo using the {\it SimProp}  package~\citep{Aloisio:2017iyh}.\footnote{We use {\it SimProp} v2r4 with the model from \cite{Gilmore:2011ks} for the extragalactic background light and the photo-disintegration cross sections from \cite{Koning:2005ezu,Koning:2012zqy} \citep[setup $M=4$, see][]{Aloisio:2013hya}.} Finally, $F(\nn,\nn_i;\Theta)$ represents the probability density function of the arrival direction $\nn$ given the direction of the galaxy $\nn_i$. We use a von Mises-Fisher distribution (equivalent to a Gaussian distribution on the sphere) centered on $\nn_i$ with an adaptable parameter $\Theta$, which is aimed at modelling in an effective way the magnetic deflections experienced by UHECRs throughout their propagation. This choice for $F$ is based on the assumption that some directionality of the sources is preserved in the observed sky maps due to the high rigidity of the particles, a reasonable assumption based on what is known on the strength of the magnetic fields experienced. A comprehensive treatment of the  deflections would require precise knowledge of the Galactic magnetic field and of the filamentary structure and strength of the Local-Sheet magnetic field, both of which are expected to imprint random-type deflections from magnetic turbulence as well as mass-spectrometric-type deflections from large-scale fields. Despite recent progresses, such a knowledge is still elusive. Our above choice allows us to emulate deflections to first order with a single model parameter, $\Theta$. Some comparisons to simulations with models of magnetic fields are further discussed in Appendix~\ref{sec:gmf}. 
 
For $z>z_\mathrm{cat}^\mathrm{max}$, the energy spectrum $J^{\mathrm{bkg}}_A(E)$ observed at present time for a given species $A$ is modelled from the background density of sources as
\begin{align}
\label{eqn:model_bkg}
    J^{\mathrm{bkg}}_A(E) &= \frac{c}{4\pi}\int \dif z \left|\frac{\dif t}{\dif z}\right| \sum_{A'} \int \dif E' \dot q_{A'}(E',z) \frac{\dif\eta_{AA'}(E,E',z)}{\dif E} \nonumber \\ 
    &= \frac{c}{4\pi} \sum_{A'} k \mathcal{E}_{A'} \int \dif z \left|\frac{\dif t}{\dif z}\right| \dot{\rho}_{\mathrm{SFR}}(z) \nonumber \\ 
    &\quad \times\int \frac{\dif E'}{E'} p_{A'}(E')\frac{\dif\eta_{AA'}(E,E',z)}{\dif E}.
\end{align}
Here, the factor $c / 4\pi$ converts the local density of UHECRs per unit energy into a flux per steradian (intensity) and per unit energy. This background contribution is assumed isotropic, a reasonable assumption beyond a few hundred Mpc \citep{Biteau:2021pru}.

\subsection{Time delays and magnetic deflections}
\label{sec:magnetic_effects}

In a transient scenario, sources are visible by an observer during a finite duration time $\Delta\tau$ only. Due to the magnetic wandering of particles en route to the Earth, the duration window depends on the magnetic rigidities of the particles, the distance from the source and the magnetic-field properties. For an observation duration much smaller than $\Delta\tau$, the mean number of relic bursts that contribute to the energy spectrum observed today on Earth reads as
\begin{equation}
\label{eqn:lambda}
    \lambda_i= S_{i} \Delta\tau.
\end{equation}
The stochastic nature of any transient scenario can thus be captured in Equation~\eqref{eqn:model_src} by randomly drawing the number of bursts in each galaxy according to a Poisson distribution of parameter $\lambda_i$. The median scenario, also known as the Asimov model, can serve as  as a benchmark for data comparison. The median $\mu_{1/2}$ of the Poisson draws is conveniently approximated here by
\begin{equation}
    \label{eqn:Poisson}
    \mu_{1/2}=\left\{
    \begin{array}{ll}
    \lambda \qquad \mathrm{for}\  \lambda > \lambda_\mathrm{th}\\
    0 \qquad \mathrm{otherwise}
    \end{array}
    \right.,
\end{equation}
with $\lambda_\mathrm{th} \simeq \ln 2$. We have verified that this approximation for the Poisson process provides the expected median over a hundred random draws of sky maps, while also considerably simplifying the computation.\footnote{The careful reader will note that this approximation violates the additive property of Poisson variables. Verification using random draws of sky maps is a necessary step in the study.}

The image on Earth of a burst of duration $\delta t$ is visible during a period $\Delta \tau \gg \delta t$, which can be shown to scale with the amplitude of the magnetic field, $B$, the particle rigidity, $R$, the source distance, $d$, and the coherence length of the field, $\lambda_B$, as~\cite[e.g.][]{1998tx19.confE.617A}
\begin{align}
\label{eqn:delta_tau}
\Delta\tau &= \frac{\sqrt{2} \lambda_B}{9c} \left(\frac{cBd}{R}\right)^2  \\
    &= 70\,\mathrm{kyr} \times \left(\frac{B}{10\,\mathrm{nG}}\right)^2\left(\frac{R}{5\,\mathrm{EV}}\right)^{-2}\left(\frac{d}{2\,\mathrm{Mpc}}\right)^2\left(\frac{\lambda_B}{10\,\mathrm{kpc}}\right).\nonumber 
\end{align}
Depending on the intervening magnetic fields, the temporal spread $\Delta\tau$ can be substantial and make a transient event quasi-persistent for an observer far-enough away from the source \cite[e.g.][for a review]{Sigl:2000vf}. The magnetic fields of interest that may enter into Equation~\eqref{eqn:delta_tau} can be found in four environments: the surrounding of the source (see Section~\ref{sec:clusters}), the extragalactic medium, the Local Sheet and the Milky Way. 

Upper limits on extragalactic magnetic fields in cosmic voids are  conservatively set to the nG level from rotation measurements~\citep{Pshirkov:2015tua,2020MNRAS.495.2607O}. Yet, they can be refined down to $10-50\,$pG for magnetic fields of primordial origin that would affect CMB anisotropies \citep{2019PhRvL.123b1301J}. On the flip side, lower limits at the fG level have also been derived from the non-observation in the GeV range of gamma-ray cascades from TeV blazars~\citep{Neronov2010,Tavecchio:2010mk,2018ApJS..237...32A, 2023ApJ...950L..16A}. Therefore, we adopt here a conservative value of 10\,pG for a coherence length of 1\,Mpc, as suggested by CMB constraints for a primordial origin of extragalactic magnetic fields and by cosmological magnetohydrodynamics simulations for an astrophysical origin \citep{2017CQGra..34w4001V}. With such characteristics, extragalactic magnetic fields in voids induce a time spread of ${\simeq}\,70\,000$ years for UHECRs at a magnetic rigidity of 5\,EV from sources at 200\,Mpc.

The Local Sheet refers to the planar structure of the cosmic web that includes the Local Group of galaxies and the ring of surrounding galaxies known as the Council of Giants. According to \cite{McCall:2014eha}, the Local Sheet can be described as a disk of diameter ${\sim}\,10$\,Mpc and thickness ${\sim}\,0.5$\,Mpc. About 1 Mpc from the centre of the sheet are Andromeda and the Milky Way. These two galaxies are at the heart of the Local Group, which can be described as a disk nearly coplanar with the Local Sheet, with a diameter of ${\sim}\,5$\,Mpc and a thickness of ${\sim}\,1.5$\,Mpc. The plane of the Milky Way is nearly perpendicular to that of the Local Sheet, as might be expected if the spin of galaxies is determined by the tidal field during the formation of the sheet through gravitational collapse \citep[see, e.g.,][and references therein]{2023MNRAS.520L..28A}. 

The magnetic field permeating the Local Sheet is largely under-constrained, as is the distribution of magnetic fields in the cosmic web. The field at the nodes of the web is constrained in the brightest galaxy clusters. The warm-hot plasma that resides in the intracluster medium embeds a turbulent magnetic field with a coherence length of ${\simeq}\,10\,$kpc and an amplitude of the order of a few $\mu $G, whose fluctuations are well represented by a Kolgomorov power spectrum \citep[e.g.][for the Coma cluster]{2010A&A...513A..30B}. The field in the structures connecting the nodes has been detected by radio and X-ray stacking of galaxy filaments \citep{2021MNRAS.505.4178V}. The authors estimate a magnetic-field amplitude in filaments of the order of 30\,nG, a value that is confirmed by Faraday rotation measurements \citep{2022MNRAS.512..945C}. The latter analysis suggests a predominantly structured field in the filaments, with a turbulent component of amplitude estimated at $3-6$\,nG. The field in the sheets between the filaments is unconstrained, both in observations and in cosmological simulations. We assume a warm-hot plasma with properties similar to those of clusters, i.e.\ a coherence length $\lambda_B \simeq 10\,$kpc. The turbulent field is presumed to have an amplitude comparable to that observed in filaments. Given the paucity of constraints on cosmic magnetism, we consider \textit{a priori} that plausible values are between 1 and 100\,nG. We adopt these values to model the Local-Sheet field as a Kolmogorov turbulence extended over the radius of the Local Group, $1-3\,$Mpc. Such radii correspond to the under-density range in Figure~\ref{fig:evolution} \citep[see also][for a detailed discussion of the size and geometry of the Local Group]{McCall:2014eha}. The geometric description of the magnetic field of the Local Sheet as a central sphere encompassing the Local Group is the simplest approximation that can be tested at this stage, since the properties of the plasma at the heart of the Local Sheet are still poorly understood both theoretically and observationally.

With a radius of 2\,Mpc and a magnetic field of $10\,$nG strength and $10\,$kpc coherence length injected into Equation~\eqref{eqn:delta_tau}, the time spread of UHECRs at a magnetic rigidity of 5\,EV accumulated in the Local Sheet amounts to ${\simeq}\,70\,000$ years for sources beyond 2\,Mpc. Consequently, the Local Sheet field is expected to show a larger contribution to $\Delta\tau$ than the extragalactic magnetic field in voids for sources within the GZK horizon. This structure also contributes to deflect particles following~\citep[see e.g.][]{1998tx19.confE.617A}
\begin{align}
\label{eqn:delta_theta}
    \Delta\theta &= \frac{2}{3} \frac{cB\sqrt{\lambda_B d}}{R} \\
     &= 10^\circ  \times\left(\frac{B}{10\,\mathrm{nG}}\right)\left(\frac{R}{5\,\mathrm{EV}}\right)^{-1}\left(\frac{d}{2\,\mathrm{Mpc}}\right)^{\nicefrac{1}{2}}\left(\frac{\lambda_B}{10\,\mathrm{kpc}}\right)^{\nicefrac{1}{2}}.\nonumber
\end{align}
The spatial extension of the Galactic magnetic field is too limited to imprint sizeable time spreads for extragalactic sources. However, with a strength of several $\mu$G in the disk, as estimated from polarization and Faraday-rotation measurements, the Galactic magnetic field is sufficiently intense to contribute to UHECR angular deflections. We account for the corresponding contributions in Appendix~\ref{sec:gmf} with the model of~\cite{Jansson2012AField}. To explore the parameter space more feasibly, we approximate the Galactic magnetic field by a Kolmogorov turbulence extending to 10\,kpc, with an amplitude of $1\,\mu$G and a coherence length of $100\,$pc. For these values, a UHECR emitted at a distance $d > 10\,$kpc with a rigidity of 5\,EV undergoes in the Milky Way a time delay of ${\simeq}\,200$\,yr and an angular deviation of ${\simeq}\,7^\circ$, which add quadratically to those induced by the Local Sheet. A summary of the default parameters used for the turbulent components of the magnetic fields relevant to this study is provided in Table~\ref{tab:b_fields}.

\begin{deluxetable}{lccc}
\tablecaption{Default parameters for turbulent magnetic fields considered in the model. \label{tab:b_fields}}
\tablehead{
\colhead{} 		& \colhead{$r_\mathrm{max}$} 	& \colhead{$B$} 	& \colhead{$\lambda_B$}\\
			& Mpc & nG & kpc}
\startdata
Milky Way		& $\phantom{0}10^{-2}$			&	$\phantom{0}10^{3\phantom{-}}$			&	$\phantom{0}10^{-1}$	\\
Local Group / Sheet	& $2$			&	$\phantom{0}10^{\phantom{-}\phantom{0}}$		&	$\phantom{0}10^{\phantom{0}\phantom{-}}$	\\
Voids			& $\infty$			&	$\phantom{0}10^{-2}$					&	$\phantom{0}10^{3\phantom{-}}$	\\
\enddata
\tablecomments{For each field, a uniform turbulence is assumed to fill a spherical region of radius $r_\mathrm{max}$ (col. 2), with an amplitude $B$  (col. 3) and coherence length $\lambda_B$ (col. 4).}
\end{deluxetable}

The effect of the delay in Equation~\eqref{eqn:delta_tau} is conveniently implemented by noting that the Equations~\eqref{eqn:lambda} and \eqref{eqn:Poisson} impose a maximum rigidity for each galaxy of ejection rate $S_i$ and distance $d_i$.\footnote{The conservation of the Lorentz factor in the photodissociation process allows the observed rigidity to be approximated by the emitted rigidity of UHECR nuclei. This approximation is limited here to the implementation of time delays, including the effect of the clusters in Section~\ref{sec:clusters}. All energy losses are otherwise fully taken into account as shown in Equations~\eqref{eqn:model_src} and~\eqref{eqn:model_bkg}.} Above this rigidity, UHECR bursts are too short and infrequent to be visible in more than half of all realizations. The median sky maps simulated above an energy threshold take into account the angular spread using a von Mises-Fisher distribution, as described in Equation~\eqref{eqn:model_src}. The angular-spread parameter is determined using Equation~\eqref{eqn:delta_theta} for $R=\bar R$, where $\bar R$ is the average rigidity of the UHECRs above the energy threshold, i.e.\ $\Theta = \Delta \theta(\bar R)$.

\subsection{Galaxy-cluster extinction}
\label{sec:clusters}

The impact of the source environment on the observed angular extent of UHECR excesses can be neglected for a distance from the source significantly greater than the size of the environment. The time delay caused by a source environment less magnetized than the Local Sheet is effectively taken into account through the threshold of the Poisson process in Equation~\eqref{eqn:Poisson}. Only an environment with a turbulent magnetic field well above the range adopted for the Local Sheet ($1-100$\,nG) could affect the simulated sky maps. This is precisely the case for galaxy clusters, which confine UHECRs long enough to act as effective calorimeters~\citep{Dolag:2008py, Kotera_2009, Harari_2016, Fang:2017zjf}. We model here the effect of galaxy clusters following \cite{Condorelli}, whose approach is summarized below. 

In \cite{Condorelli}, the hot, tenuous gas of the intracluster medium is described under the self-similar assumption for each galaxy cluster, so that its radial pressure profile can be described using only its mass and redshift \citep{KAISER_1986}. The magnetic field is scaled to the thermal components of the intracluster medium, assuming that magnetic energy density is proportional to thermal energy density, with a relation calibrated using the Perseus cluster \citep[$(l, b) = (150.6^\circ, -13.3^\circ)$, $d\simeq 70\,$Mpc, $M \simeq 5.8 \times 10^{14}\, M_\odot$, $R_{500} \simeq 1.3\,$Mpc, from][]{2014MNRAS.437.3939U}. With these ingredients in hand, the propagation of UHECRs along the gas-density and magnetic-field profiles can be simulated inside any galaxy cluster. A modified version of {\it SimProp}, which includes hadronic processes and magnetic-field confinement, has enabled \cite{Condorelli} to construct generic transparency functions of clusters as a function of UHECR rigidity for protons on the one hand, and nuclei on the other. 

The brightest X-ray galaxy cluster in the GZK horizon is the Virgo cluster \citep[$(l, b) = (279.7^\circ, 74.5^\circ)$, $d \simeq 15\,$Mpc, $M \simeq 1.2 \times 10^{14}\, M_\odot$, $R_{500} \simeq 0.8\,$Mpc, from][]{2016A&A...596A.101P}, followed by the Perseus cluster which is $1.5-2$ times fainter in X-rays. These clusters are opaque to both nuclei and protons in the rigidity range of interest. We apply the transparency function of \cite{Condorelli} to all galaxies behind disks of radius $3R_{500}$ centered on Virgo and Perseus. The next cluster is Coma, which is ${\simeq}\,5$ times fainter than Virgo in X-rays. We have verified that implementing the UHECR opacity of Coma or fainter clusters has no impact on the reconstructed sky maps in the energy range of interest. We do note, however, that mass composition anisotropies might be expected in the energy range where protons and heavier nuclei mix (near the ankle), due to the different filtering that these species undergo in the traversed intracluster media. We leave the exploration of the UHECR sky around the ankle to future studies that may elaborate on this framework.

\section{Constraints on burst rate density from UHECR observations}
\label{sec:results}

\begin{figure*}[t]
\centering
\includegraphics[width=0.49\textwidth]{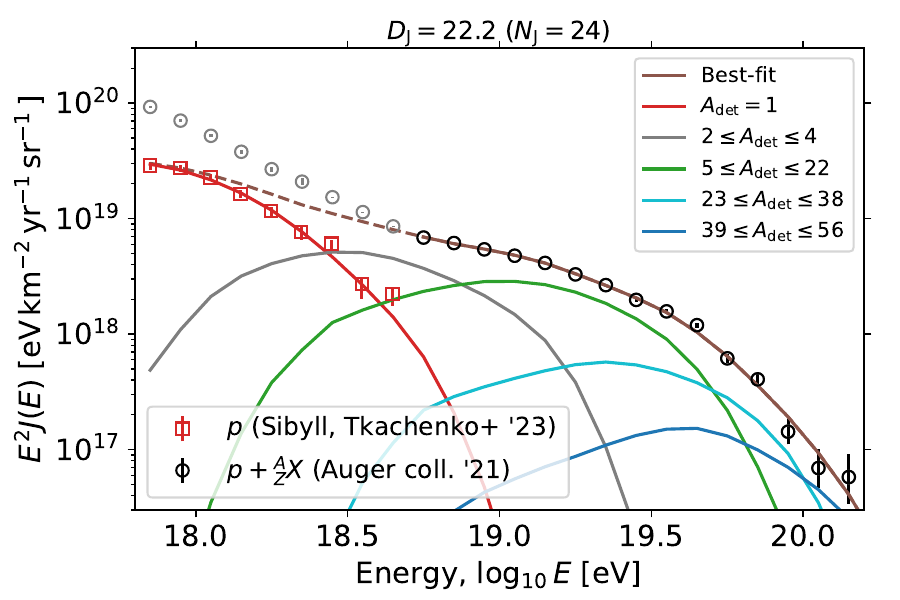}
\includegraphics[width=0.49\textwidth]{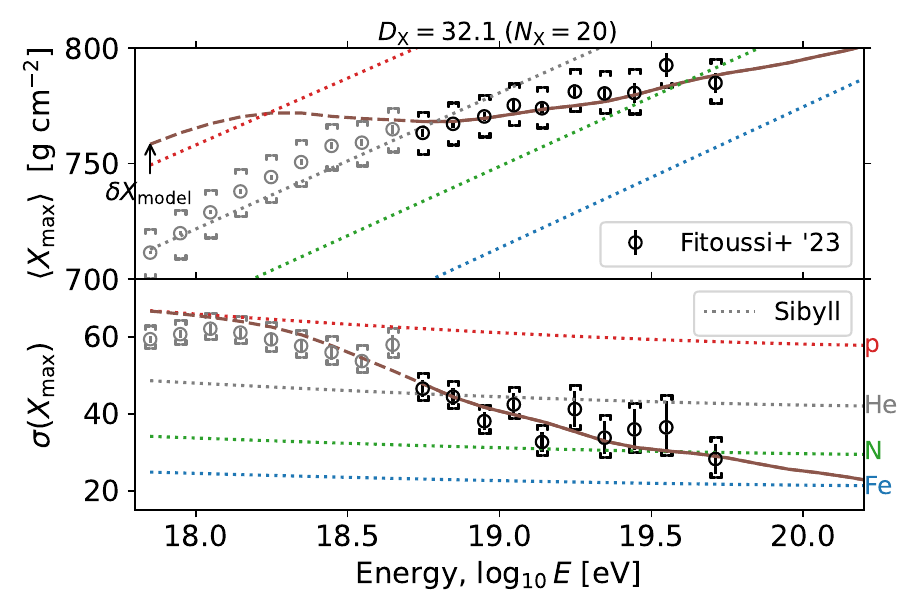}
\caption{The energy spectrum (\textit{left}) and $X_\mathrm{max}$ moments (\textit{right}) of UHECRs as a function of energy. The data points are from the latest datasets of the Pierre Auger Observatory \citep{PierreAuger:2021hun, AbdulHalim:20239}, with a proton fraction based on the Sibyll 2.3d hadronic interaction model \citep[see][]{PierreAuger:2023xfc}. The best-fit model, including a shift in $X_\mathrm{max}$ scale by $\delta X_\mathrm{model} =(1.20\pm0.18) \times \sigma_\mathrm{sys.} = 9.6 \pm 1.4$\,g\,cm$^{-2}$, is shown as a brown solid line in its range of applicability, and as a brown dashed line at lower energies.}
\label{fig:spec_xmax_bestfit}
\end{figure*}

\subsection{Best fit of spectrum and mass-composition data} 
\label{sec:spectrum_compo_bestfit}

The spectrum and mass-composition data used in this work are those obtained at the Pierre Auger Observatory \citep{PierreAuger:2015eyc}, which provides the largest exposure to date to UHECRs. The Observatory is a hybrid system that detects the extensive air showers induced by the collisions of UHECRs with atmospheric molecules. We use the all-particle energy spectrum inferred from these data~\citep{PierreAuger:2021hun} and a proxy of the primary mass of the particles~\citep{PierreAuger:2014gko,AbdulHalim:20239}, i.e.\ the slant depth of maximum of shower development, $X_{\mathrm{max}}$. Using hadronic-interaction generators to model the development of the showers, the $X_{\mathrm{max}}$ distributions allow the energy-dependent mass composition to be deduced on a statistical basis. As in \cite{PierreAuger:2023xfc}, we used the two up-to-date hadronic-interaction generators that best describe the $X_{\mathrm{max}}$ distributions, namely EPOS-LHC~\citep{Pierog:2013ria} and Sibyll2.3d~\citep{2020PhRvD.102f3002R}. We adopt the latter model in our reference configuration.

The best fit of spectrum and mass-composition data is obtained following \cite{Luce:2022awd}, with a source evolution following either the SFRD or the SMD shown in Figure~\ref{fig:evolution}. With more details than given in Section~\ref{sec:source_model}, we generically model the ejection rate per comoving unit volume and per unit energy of nucleons as
\begin{equation}
\label{eqn:qN}
    \dot Q_{\rm p}(E) = \dot Q^0_{{\rm p}}\left(\frac{E}{Z_{\rm p} E_0}\right)^{-\gamma_{\rm p}}f_{\mathrm{supp}}(E,Z_{\rm p}),
\end{equation}
with $Z_{\rm p}=1$, that is a single reference ejection rate $\dot Q^0_{{\rm p}}$, spectral index $\gamma_{\rm p}$ and suppression function $f_{\mathrm{supp}}(E,Z_{\rm p})$ for both escaping protons and protons from neutron decay. Here, $E_0$ is arbitrarily set to 1\:EeV. The ejection rate of nuclei with mass number ${A}_i$ is also generically modeled as
\begin{equation}
\label{eqn:qAapp}
    \dot Q_{{A}_i}(E) = \dot Q^0_{{A}_i}\left(\frac{E}{Z_{{A}_i} E_0}\right)^{-\gamma_{A}}f_{\mathrm{supp}}(E,Z_{{A}_i}),
\end{equation}
that is a single spectral index $\gamma_{A}$ and four independent reference ejection rates $\dot Q^0_{{A}_i}$ for helium, nitrogen, silicon and iron. The suppression function adopted both for nucleons and nuclei is taken as
\begin{equation}
\label{eqn:fsupp}
    f_{\mathrm{supp}}(E,Z_A) = 
    \begin{cases}
    1 & \mathrm{if~}E\leq E^{Z_A}_{\mathrm{max}},\\
    \exp{\left(1-\nicefrac{E}{E^{Z}_{\mathrm{max}}}\right)} & \mathrm{otherwise}.
    \end{cases}
\end{equation}
The maximum acceleration energy is assumed to be proportional to the electric charge of each nucleus, $E^{Z_A}_{\mathrm{max}} = Z_A E_{\mathrm{max}}$, with a single free parameter $E_{\mathrm{max}}$ shared by all species. To connect properties at the sources with observables at Earth, the extragalactic propagation is carried out with \textit{SimProp}. All the events generated are stored in a normalized 5D array $T\left(E_{\rm det}, A_{\rm det} \middle|z, E_{\rm inj}, A_{\rm inj}\right)$. In this way, the tensor encodes the probability that a particle with energy $E_{\rm inj}$ and mass $A_{\rm inj}$ emitted at a redshift $z$ is detected with an energy $E_{\rm det}$ and a mass $A_{\rm det}$. After taking into account the possible interactions of UHECRs with photons of the CMB and of the extragalactic background light during the extragalactic propagation, the flux at Earth can be calculated and then compared to the experimental data. This comparison is made using a deviance, $D$, similar to that presented in Equation~(6) of \cite{Luce:2022awd}. The first deviance term is defined here as a $\chi^2$ including the points of the proton spectrum at energies below the ankle, the all-particle spectrum above the ankle, and the mean and standard deviation of $X_\mathrm{max}$ above the ankle (red and black markers in Figure~\ref{fig:spec_xmax_bestfit}). The second deviance term accounts for the systematic uncertainty on the mean $X_\mathrm{max}$ values, $\sigma_\mathrm{sys.} = 8\,$g\,cm$^{-2}$, so that \ $D = \chi^2 + \left(\delta X_\mathrm{model}/\sigma_\mathrm{sys.}\right)^2$, where $\delta X_\mathrm{model}$ is an energy-independent shift applied to the model.

We present the results obtained with a source evolution following the SFRD in Figure~\ref{fig:spec_xmax_bestfit}. Consistent to that found in other studies, the value of $E_{\mathrm{max}}$ is determined primarily by the drop in the proton component at $E_{\mathrm{max}} = (1.9 \pm 0.1)\,$EeV. The spectral index of the nuclei, $\gamma_{A}= -0.36  \pm  0.21$, is in turn determined by the increase of the average mass with energy, which is almost mono-elemental. The best-fit solution therefore imposes a hard index for nuclei so that the contribution of each element mixes as little as possible. Protons are present in an energy range where a mixture of elements is required to model the all-particle spectrum and the composition. The best-fit value of $\gamma_{\mathrm{p}}= 2.6  \pm  0.7$ is much softer than that of $\gamma_{\mathrm{A}}$. The spectral index of the proton component is affected by a large uncertainty. Indeed, in a scenario where the density of sources is larger in the past, a non-negligible fraction of the protons observed today originate from the interaction of heavier nuclei during their extragalactic propagation. Thus, for a total emissivity per unit SFR of $(4.8 \pm 0. 2) \times 10^{46}$ erg\,$M_\odot^{-1}$, only $(19 \pm 4)\,\%$ of the energy is emitted as protons, while the proportions of energy emitted as He, CNO, Si and Fe are $(14 \pm 3)\,\%$, $(48 \pm 4)\,\%$, $(15 \pm 3)\,\%$ and $(4 \pm 1)\,\%$. 

Note that the deviance of the fit, $D=54.3$ for 36 degrees of freedom, is significantly improved by shifting the $X_{\mathrm{max}}$ scale of the model for UHECR protons nuclei by $\delta X_\mathrm{model} =(1.20\pm0.18) \times\sigma_\mathrm{sys.}$. This would correspond to a data shift by $-(1.20 \pm 0.18)\times\sigma_\mathrm{sys.}$ which is in line with values ranging from $-0.9\times\sigma_\mathrm{sys.}$ to $-0.5\times\sigma_\mathrm{sys.}$ as identified already in \cite{PierreAuger:2016use}. Systematic uncertainties in the $X_\mathrm{max}$ scale, $\delta X_\mathrm{model}$, should also impact the proton spectrum below the ankle, but this effect is difficult to take into account with the public data that are provided for a zero value of this nuisance parameter only. The systematic uncertainty of $\sigma_E/E \simeq 14\,\%$ in the energy scale has a significant impact on the all-particle and proton spectra. A shift of the model by $+1\,\sigma_E$ leads to a lower emissivity $(2.4 \pm 0.2) \times 10^{46}\,$erg\,$M_\odot^{-1}$ and a better deviance ($D = 46.0$). A shift by $-1\,\sigma_E$ leads to higher emissivity $(9.7 \pm 0.4) \times 10^{46}\,$erg\,$M_\odot^{-1}$ and a worse deviance ($D = 61.4$). 

For a source evolution following the SMD, the emissivity per unit $M_\star$ is estimated to $(4.2 \pm 0.2) \times 10^{36}\,$erg\,$M_\odot^{-1}$\,yr$^{-1}$, with a cutoff energy at $2.6 \pm 0.1$\,EeV, a proton index of $3.0 \pm 0.2$ and a nuclear index of $0.6 \pm 0.1$. The emitted proton component, which saturates the observed fraction, carries $(54 \pm 4)\, \,\%$ of the energy, $(32 \pm 4)\,\%$ being carried by CNO and less than 5\,\% carried by each of the other species. The deviance in the SMD scenario reaches $68.3$ units for a shift in the $X_\mathrm{max}$ scale of the model by $(1.5 \pm 0.2)\times\sigma_\mathrm{sys.}$. These fits of the model to the data can be easily reproduced using the public version of the analysis code, which we make available \citep[see][]{condorelli_2024_11440864}.

\subsection{Full-sky observations above ${\simeq}\,40\,$EeV}
\label{sec:augerta_maps}

\begin{figure*}[ht!]
\centering
\includegraphics[width=0.7\textwidth]{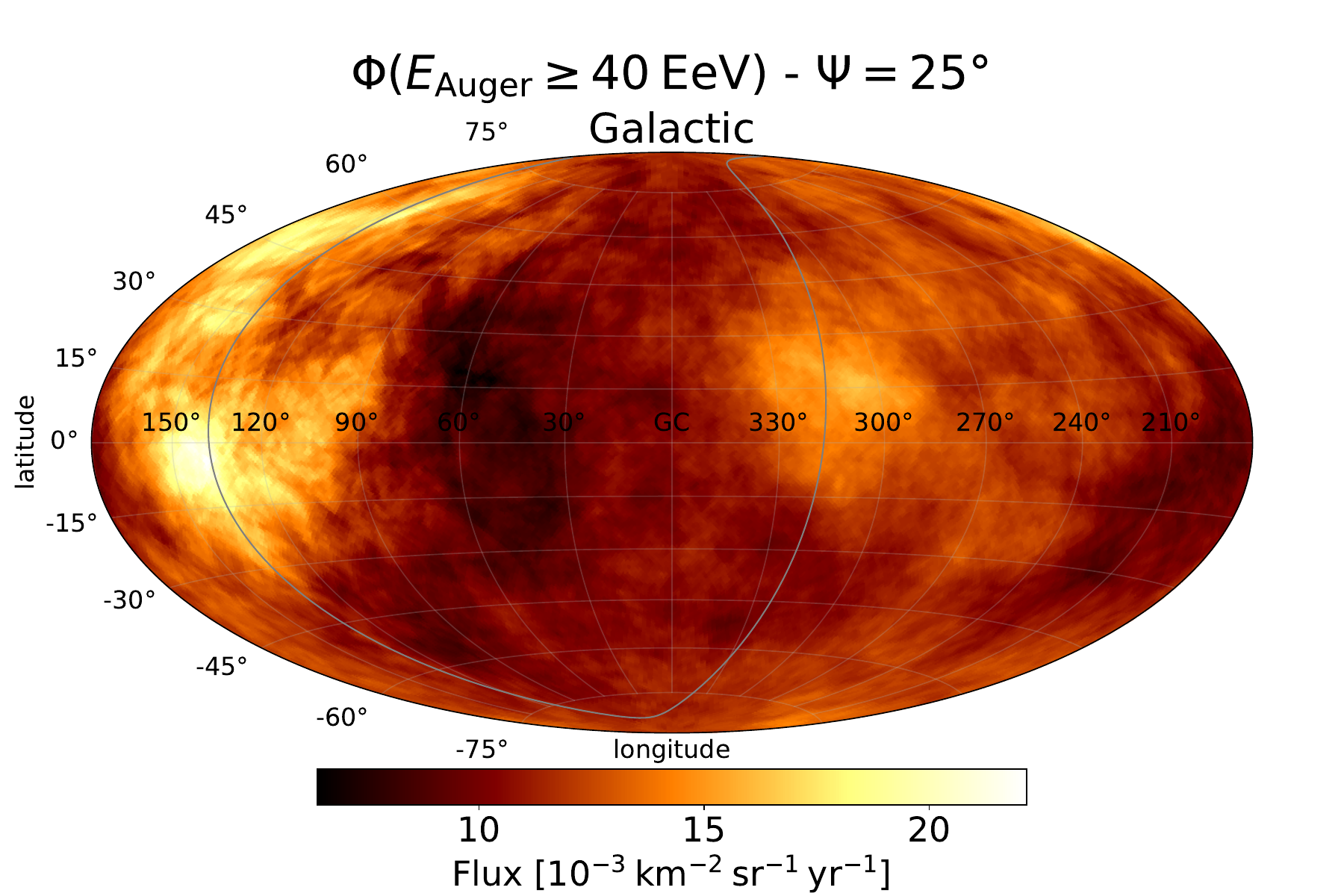}
\caption{Flux map of UHECRs at energies above 38\,EeV (48\,EeV) as reconstructed from the Pierre Auger Observatory (Telescope Array). The sky map is smoothed using a top-hat function of radius $\Psi = 25\,^\circ$. The grey line shows the supergalactic plane. Adapted from \cite{PierreAuger:2023mvf}.}
\label{fig:augerta}
\end{figure*}

The Pierre Auger Observatory, covering 3\,000\,km$^2$ in the province of Mendoza (Argentina), has since 2004 accumulated a total exposure of ${\simeq}\, 135\,000\,$km$^2$\,sr\,yr over nearly the entire UHECR sky, except in the northernmost quarter. In parallel, the Telescope Array~\citep{TelescopeArray:2012uws}, covering 700\,km$^2$ in Utah (USA), has provided since 2008 a view of the northern UHECR sky with a total exposure of ${\simeq}\,17\,500\,$km$^2$\,sr\,yr. A series of inter-collaborative efforts has resulted in the full-sky map of arrival directions shown in Figure~\ref{fig:augerta} in Galactic coordinates \citep{PierreAuger:2023mvf}, with a top-hat smoothing on an angular scale $\Psi = 25^\circ$ \citep[corresponding to a Gaussian angular scale $\Theta_\mathrm{obs} \approx \Psi/1.59 \simeq 16\,^\circ$, see discussion in][]{PierreAuger:2022axr}. Although the techniques for assigning energies to events are nearly the same at both observatories, there are differences in the way primary energies are derived. To produce the sky map, the energy estimators are cross-calibrated by ensuring that the intensities observed at each observatory are identical in the area common to both fields of view. On account of this, the sky map is built with events at energies above 38\,EeV (48\,EeV) in terms of the energy scale of the Pierre Auger Observatory (Telescope Array). For convenience, we choose to refer to ${\simeq}\,40$\,EeV for the energy threshold hereafter. Note that, as accumulated exposure in the Northern Hemisphere is around eight times lower than in the Southern Hemisphere, the fluxes reached by the excesses are prone to greater statistical uncertainties on the northernmost quarter of the sky, i.e.\ on the left quarter of the map in Galactic coordinates in Figure~\ref{fig:augerta}. These uncertainties are illustrated by the Poisson fluctuations visible at an angular scale of ${\simeq}\,5^\circ$, which corresponds to the typical spacing between each of the ${\sim}\,2\,000$ UHECRs detected over the entire sphere at $E \gtrsim 40\,$EeV.

Due to the GZK effect, the isotropic background of UHECR arrival directions caused by sources distributed throughout the distant universe is suppressed. The overdensities hinted at in Figure~\ref{fig:augerta} could therefore point to dominating foreground sources. However, a blind search for excesses does not reveal any statistically significant indications for anisotropy \citep{PierreAuger:2020mkh, PierreAuger:2022axr}. Indeed, by not specifying \textit{a priori} the targeted regions of the sky where the excesses are searched for, nor the angular-window radius nor the energy threshold, the analysis sensitivity suffers from the numerous  trials performed.

Yet, even without compelling proof for discrete sources, a correlation between UHECR arrival directions and the flux pattern of a class of astrophysical objects might reveal earlier an anisotropy that would trace the sources. The best evidence is currently provided by a catalog of bright star-forming or starburst galaxies \citep{PierreAuger:2018qvk,PierreAuger:2022axr} with a confidence level of $4.6\,\sigma$ when considering full-sky maps \citep{PierreAuger:2023mvf}. The analysis assesses the clustering of UHECR arrival directions around the predefined astrophysical source positions, in proportion to the non-thermal photon luminosity of the host galaxies. The similarities between the model and observations, best captured on an angular scale of $\Theta_\mathrm{obs} = (15_{-3}^{+5})^\circ$, are due to overdensities in the direction of Local-Sheet galaxies. However, the low signal strength,  $(12_{–3}^{+5})\,\%$, precludes the conclusion that UHECRs come exclusively from these bright nearby galaxies. 

Interestingly, the brightest star-forming galaxies known as starburst galaxies turn out to be responsible for ${\simeq}\,15\,\%$ of the total SFR for redshifts $z<2$ \citep{Rodighiero:2011px,Sargent:2012rj}. If UHECR sources are traced by the SFR (or $M_\star$), then the signal intensity could suggest that the sources are hosted by every star-forming galaxy in the universe. With their higher SFR per unit $M_\star$, starburst galaxies would merely have a higher probability of hosting UHECR transient sources than main-sequence galaxies with similar $M_\star$. The sky maps constructed below allow us to explore the rate of bursts in each galaxy relative to its SFR (or $M_\star$) that could, given the observed angular scale of the correlation, accommodate the observed signal strength. 

\subsection{Constraints on $k$ from arrival directions}
\label{sec:median_maps}

As shown in Equation~\eqref{eqn:lambda}, the number of bursts in each galaxy can be drawn at random according to a Poisson process of parameter $\lambda_i$. For a set of magnetic fields, the Poisson parameter is governed by the rate of bursts $k$ relative to the SFR (or $M_\star$) of the host galaxy (see Equation~\eqref{eqn:Si}). For a set of parameters fitted to the energy spectrum and mass composition data (see Section~\ref{sec:spectrum_compo_bestfit}), the intensity of UHECRs observed on Earth in each direction above ${\simeq}\,40\,$EeV is then estimated from Equations~\eqref{eqn:model_src} and~\eqref{eqn:model_bkg}. The results are illustrated, for a given $k$ value, by means of the median intensity in each direction. These median sky maps better reflect the stochastic nature of the transient scenario than the mean sky maps, which would just depict the results obtained by weighting each galaxy by $\lambda_i$. 

The value of $k$ affects the sky maps as follows. For $k$ values such that the emission rates are large ($\lambda_i > 1$), each galaxy emits almost persistently. The resulting burst rate is large enough to imply dominant contributions from galaxies of the Local Group, in particular Andromeda ($d \simeq 750\,$kpc), the Magellanic clouds ($d \simeq 50-60\,$kpc) and the Milky Way. For lower values of $k$, galaxies in the Local Group are filtered out. This is because the time spread induced by the Local-Sheet magnetic field is not sufficient to make the contribution of nearby galaxies stationary: their burst rate is so low that the observer is likely to be between two bursts. For even lower values of $k$, the dimmest galaxies in the Local Sheet and beyond disappear due to their low burst rate.

The transient emission from our Galaxy requires special treatment due to its proximity. The sky maps can be evaluated by concentrating all Galactic matter in its bulge, centered on Sgr A*, using the SFR and $M_\star$ from \cite{2015ApJ...806...96L} normalized to the Chabrier initial mass function. Alternatively, the sky maps can be evaluated omitting the matter contained in the Milky Way, which yields more pessimistic constraints. These assumptions encompass the results expected for a three-dimensional distribution of Galactic matter, the treatment of which within a transient model goes beyond the scope of this work.

We explore the range of $k$ values compatible with the data using the two most prominent excesses identified by the Pierre Auger and Telescope Array collaborations, i.e.\ the excesses in the Centaurus (Cen) and Ursa Major (UMa) regions, around the directions $(l,b) = (305^\circ, 16^\circ)$ and  $(l,b) = (176^\circ, 45^\circ)$ in Galactic coordinates. These two excesses are responsible for the evidence of anisotropy above ${\simeq}\,40\,$EeV \citep{PierreAuger:2023mvf}. The implications of our model for a potential  excess in the Andromeda region (eponymous galaxy located at $(l,b) = (121.2^\circ, -21.6^\circ)$) will be discussed.

For a given value of the Local-Sheet turbulent magnetic field and of the burst rate, we build the median sky map and search for the local maximum of the flux within a radius of $40^\circ$ around the Cen excess and that around the UMa excess, as localized by the Pierre Auger and Telescope Array collaborations, respectively. The relatively large value of the search radius ($40^\circ$) effectively takes into account the displacement of maxima expected as a result of coherent magnetic fields, but whose exact direction remains dependent on the model \citep{2016APh....85...54E}. If any of the local maxima falls at the edge of its search region, the model is considered to lack one of the hotspots hinted at by the data, and the parameter set is rejected. 

\begin{figure}[t!]
\centering
\includegraphics[width=\columnwidth]{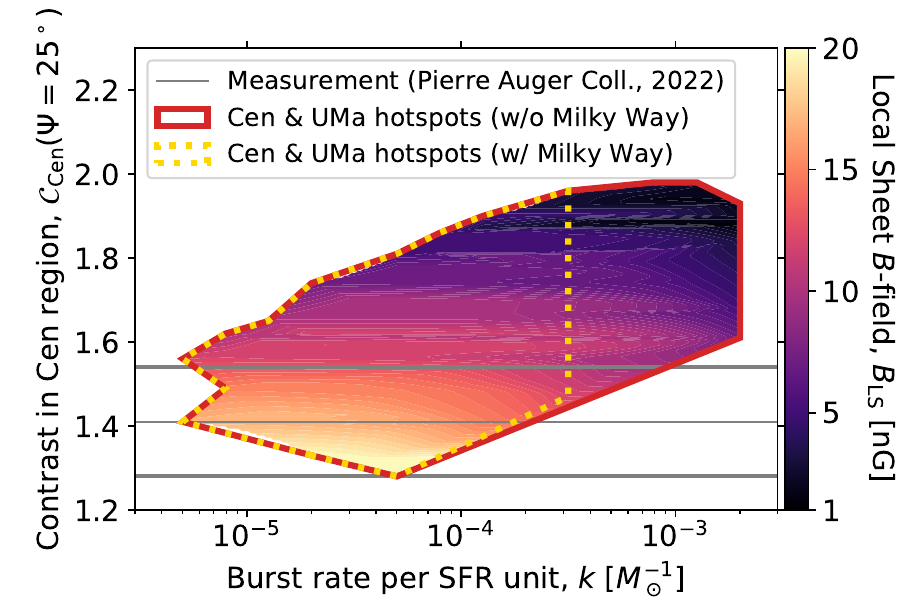}
\caption{Contrast between maximum brightness in the Centaurus region and the Southern Hemisphere average. The 68\,\% confidence region from the Pierre Auger Observatory is shown with grey lines \citep{PierreAuger:2022axr}. The parameter space is explored with a scan of the burst rate, $k$ ($x$-axis), and of the magnetic field in the Local Sheet, $B_\mathrm{LS}$ (color bar). The solid red line delimits the parameter space for which localized excesses are observed in the Centaurus (Cen) and Ursa Major (UMa) regions, for a conservative approach excluding the Milky Way from the model. The inclusion of transient emission from our Galaxy in the model, as a point source co-localized with Sgr A*, leads to the allowed region delimited by the gold dotted line.}
\label{fig:scan_k}
\end{figure}

The parameter space allowed for a UHECR emission traced by the SFR is shown in Figure~\ref{fig:scan_k}. In an SFR ($M_\star$) model, the magnetic field of the Local Sheet is between 1 and 20 nG ($0.5-10$ nG) for a burst rate $k \in [8\times10^{-6}; 2\times10^{-3}]\,M_\odot^{-1}$ ($\dot k \in [6\times10^{-16}; 4\times10^{-13}]\,M_\odot^{-1}$yr$^{-1}$). Including the Milky Way in the model brings the upper bound of the allowed rate range down to $k \leq 3 \times 10^{-4}\, M_\odot^{-1}$ ($\dot k \leq 2\times10^{-14}\, M_\odot^{-1}$yr$^{-1}$). Added to the effect of the Galactic magnetic field, the deflections induced by a 10\,nG field in the Local Sheet lead to an angular spread $\Theta \approx 18^\circ$, in agreement with observations \citep[$\Theta_\mathrm{obs} = (15_{-3}^{+5})^\circ$,][]{PierreAuger:2023mvf}.

\begin{figure*}[ht!]
\centering
\includegraphics[width=0.7\textwidth]{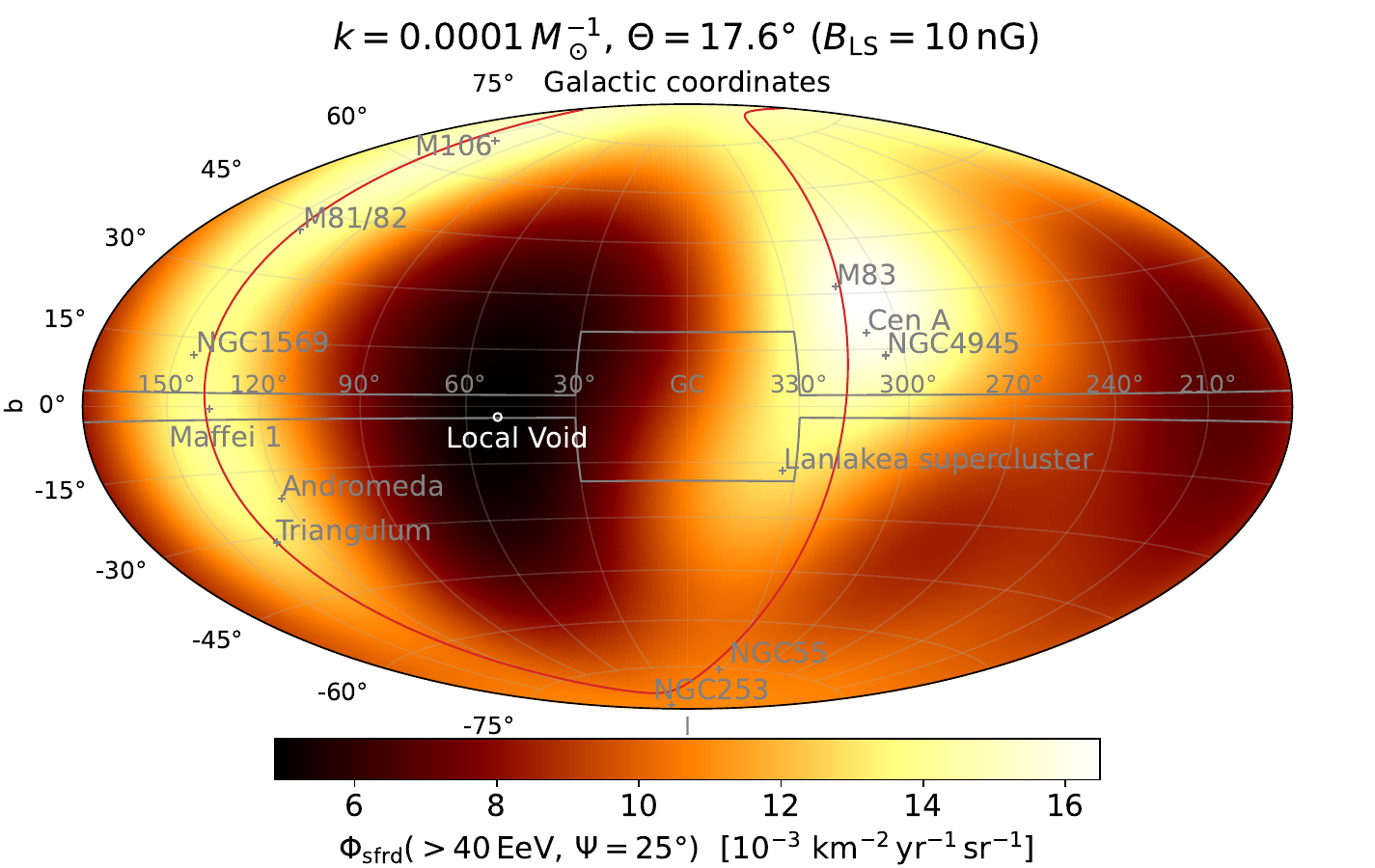}
\caption{Model flux map of UHECRs at energies above 40\,EeV for a burst rate per unit SFR $k = 1\times10^{-4}\,M_\odot^{-1}$. The sky map is smoothed using a top-hat function of radius $\Psi = 25\,^\circ$. The red line shows the supergalactic plane. The area delimited by the grey line represents the zone of avoidance. Following \cite{Biteau:2021pru}, prominent galaxies and structures contributing to the flux are represented by grey markers and the center of the Local Void is indicated by a white circle. The complete figure set (7 images) is available in  Appendix~\ref{sec:arXiv}.}
\label{fig:modelmaps}

\end{figure*}

In the suppression region of the UHECR spectrum, given the steep decrease of the number of events with energy, the flux value derived for each hotspot is highly dependent on the energy threshold considered. Rather than the flux, we provide instead the contrast for the hotspot measured with the best accuracy. Data from the Pierre Auger Observatory yield a Cen hotspot flux reaching $\Phi_\mathrm{Cen}({>}\,40\,\mathrm{EeV}) =(15.9 \pm 1.3) \times 10^{-3}\,$km$^{-2}\,$yr$^{-1}\,$sr$^{-1}$ in a top-hat region of radius $\Psi = 25^\circ$ \citep{PierreAuger:2022axr}. This maximum flux reconstructed by the Observatory can be compared with the average value obtained in the Southern Hemisphere, $\Phi_\mathrm{SH}({>}\,40\,\mathrm{EeV}) = (11.3 \pm 0.4) \times 10^{-3}\,$km$^{-2}\,$yr$^{-1}\,$sr$^{-1}$, at declinations $\delta < 45^\circ$. The 68\,\% confidence region on the Cen contrast, $\mathcal{C}_\mathrm{Cen} = \Phi_\mathrm{Cen} / \Phi_\mathrm{SH} = 1.41 \pm 0.13$, is shown with grey lines in Figure~\ref{fig:scan_k}. As illustrated by the colored region in Figure~\ref{fig:scan_k}, the SFR scenario predicts contrast values in $1.3-2$ that are compatible with observations in the Southern Hemisphere. A somewhat narrower range of contrast values, $1.5-1.85$, is obtained in the $M_\star$ scenario, which is also compatible with observations. In both scenarios, comparable contrasts are predicted for the UMa and Cen hotspots. Despite the good agreement between expected and observed contrasts, we do not use this observable to constrain the parameter space, either for the magnetic field in the Local Sheet or for the UHECR burst rate. The flux in these regions is indeed subject to the effects of deflection and (de)magnification by the coherent components of the traversed magnetic fields (see Appendix~\ref{sec:gmf}).

The contrast obtained for the third hotspot in the Andromeda region (see Figure~\ref{fig:modelmaps}, where the location of the eponymous galaxy is marked) depends strongly on the region of parameter space considered. This statement is illustrated in Appendix~\ref{sec:arXiv}, where the sky model is explored for different values of the burst rate and for a fixed amplitude of the turbulent magnetic field in the Local Sheet. For the highest burst rates ($k\geq 2 \times 10^{-4}\,M_\odot^{-1}$), the Andromeda and Triangulum galaxies produce an excess brighter than those from the Local-Sheet galaxies and from our supercluster Laniakea. For $k = 1 \times 10^{-4}\,M_\odot^{-1}$, illustrated by the model map in Figure~\ref{fig:modelmaps} that appears to match the observed map in Figure~\ref{fig:augerta}, the catalog up to $d_\mathrm{cat}^\mathrm{max} = 350\,$Mpc accounts for ${\simeq}\,97\,\%$ of the flux above 40\,EeV, with only ${\simeq}\,3\,\%$ of the flux coming from the sphere of radius $1$\,Mpc that includes Andromeda and Triangulum, ${\simeq}\,20\,\%$ from the $1-10$\,Mpc shell that includes the galaxies of the Council of Giants, and ${\simeq}\,50\,\%$ from the shell between $10-100$\,Mpc that contains our supercluster Laniakea. Laniakea contributes about two thirds of the Cen peak excess, while the peak emission from the Perseus-Pisces supercluster is almost negligible.

In summary, by simply requiring the presence of hotspots in the UMa and Cen regions, the model accurately reproduces the patterns of flux over- and under-density across the sky as seen by the Pierre Auger and Telescope Array observatories. The model in Figure~\ref{fig:modelmaps} also naturally reproduces the amplitude of the flux excesses in the hotspot regions, as well as the angular correlation scale, $\Theta_\mathrm{obs}$, for the magnetic-field parameters in the Local Sheet in Equation~\eqref{eqn:delta_theta}.

\newpage
\section{Astrophysical sources of UHECRs}
\label{sec:discussion}

\subsection{Transient rather than persistent sources}

We have hypothesized that the sources responsible for the acceleration of UHECRs and for the localized excesses in the sky are transient. This hypothesis stems from a criterion by \cite{1984ARA&A..22..425H}, which requires UHECRs to remain confined for at least one gyration within the emitter. The Hillas criterion can be neatly reformulated by noting that the vacuum impedance $\Omega_\mathrm{vac} = (\epsilon_0 c)^{-1} = 377\,\Omega$ imposes losses at the source of the order of \citep{1976Natur.262..649L, 1995PhRvL..75..386W, 2000PhST...85..191B}
\begin{align}
\label{eqn:Hillas_Lovelace}
L_B &> \frac{\Gamma^2}{\beta}  \frac{R^2}{\Omega_\mathrm{vac}}\nonumber \\
	&> 6.6 \cdot 10^{43}\, \mathrm{erg\,s}^{-1} \times \left(\frac{\Gamma^2/\beta}{100}\right)\left(\frac{R}{5\,\mathrm{EV}}\right)^2,
\end{align}
where $L_B$ is the bolometric luminosity and $(\Gamma, \Gamma \beta)$ is the four-velocity of the ejecta within which acceleration takes place. Maintaining such a non-thermal luminosity over a long period of time is difficult. For example, the magnetized winds of the starburst galaxy M\,82 carry only $5 \times 10^{41}\, \mathrm{erg\,s}^{-1}$ \citep{2013A&A...549A.118C}. Even the bolometric emission of M\,82, $2.5 \times 10^{44}\, \mathrm{erg\,s}^{-1}$ essentially of thermal origin, only slightly exceeds the Lovelace-Waxman-Blandford limit in Equation~\eqref{eqn:Hillas_Lovelace}. 

The evidence of anisotropy correlated with galaxies in the Local Sheet could come instead from the echo of a past burst from the jetted active galactic nucleus of Centaurus\,A \citep[deboosted non-thermal bolometric luminosity of $10^{43}\, \mathrm{erg\,s}^{-1}$,][]{2021MNRAS.500.3536B}. The excesses of UHECRs in the sky would then be due to reflections of the past burst off the magnetized winds of M\,82 and of the other starburst galaxies in the Local Sheet \citep{2022MNRAS.511..448B}. For the viability of this scenario, heavy nuclei must be injected into the jets of Centaurus\,A and other active galactic nuclei \citep[see e.g.\ the ``espresso'' model of][]{2021ApJ...921...85M}. Whether such scenarios meet the production rate requirements remains to be determined. The alternative that can be investigated in a more quantitative way is a transient origin of UHECRs from stellar explosions within all star-forming galaxies, as discussed here.

\subsection{Stellar rather than galactic size}

To evaluate the viability of any astrophysical scenario involving ultra-high-energy nuclear emission, the metallicity of the accelerated ejecta must be examined. The primary or secondary origin of the protons detected around the ankle of the UHECR energy spectrum depends on the evolution of the injection with redshift (see Section~\ref{sec:spectrum_compo_bestfit}). Thus, the metallicity of the source ejecta, i.e.\ the ratio of nuclei heavier than helium to protons, is poorly constrained. The ratios of helium to metals and of heavy to intermediate-mass nuclei prove to be more constraining observables. 

Because of the very hard spectrum of the nuclei, of index $\gamma_A < 1$, the integral of the emitted spectrum is convergent: the number of emitted nuclei of each species is proportional to $\mathcal{E}_A/E_\mathrm{max}^{Z_A}$. The mass fractions of  He to metals is estimated from the best-fit parameters provided in Section~\ref{sec:spectrum_compo_bestfit} as
\begin{align}
    \label{eq:abundance_ratio2}
    \left.\frac{M(\mathrm{He})}{M(\mathrm{C-Fe})}\right|_\mathrm{UHECR} &= \frac{A_\mathrm{He}\frac{\mathcal{E}_\mathrm{He}}{E_\mathrm{max}^{Z_\mathrm{He}}}}{A_\mathrm{CNO}\frac{\mathcal{E}_\mathrm{CNO}}{E_\mathrm{max}^{Z_\mathrm{CNO}}}+A_\mathrm{Si}\frac{\mathcal{E}_\mathrm{Si}}{E_\mathrm{max}^{Z_\mathrm{Si}}}+A_\mathrm{Fe}\frac{\mathcal{E}_\mathrm{Fe}}{E_\mathrm{max}^{Z_\mathrm{Fe}}}} \nonumber \\   
    & = \frac{\mathcal{E}_\mathrm{He}}{\mathcal{E}_\mathrm{CNO}+\mathcal{E}_\mathrm{Si}+\mathcal{E}_\mathrm{Fe}\frac{A_\mathrm{Fe}}{2Z_\mathrm{Fe}}} \nonumber \\   
    & = 0.21 \pm 0.05_\mathrm{stat.} \pm 0.06_\mathrm{sys.},
\end{align}
where the first uncertainty, being statistical, takes into account the covariance between the nuclear components and where the second, being systematic, characterizes the variations in energy scale, hadronic-interaction generator and redshift evolution. The value inferred for UHECRs is two orders of magnitude below $\left.\nicefrac{M(\mathrm{He})}{M(\mathrm{C-Fe})}\right|_\mathrm{ISM} =  18 \pm 2$ \citep{2009LanB...4B..712L}, which rules out an ejecta drawn from the interstellar medium.

\begin{figure*}[ht!]
\centering
\includegraphics[width=\columnwidth]{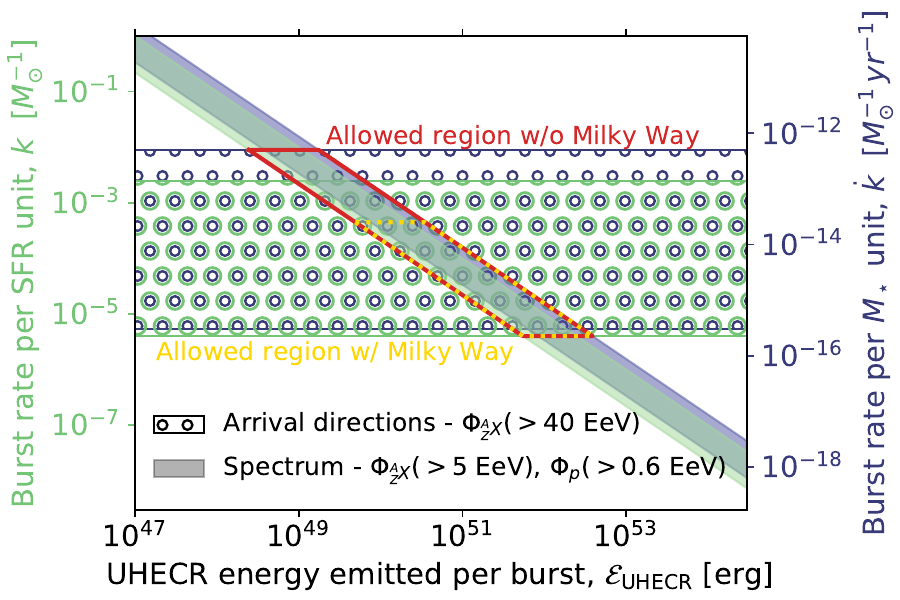}\hfill\includegraphics[width=\columnwidth]{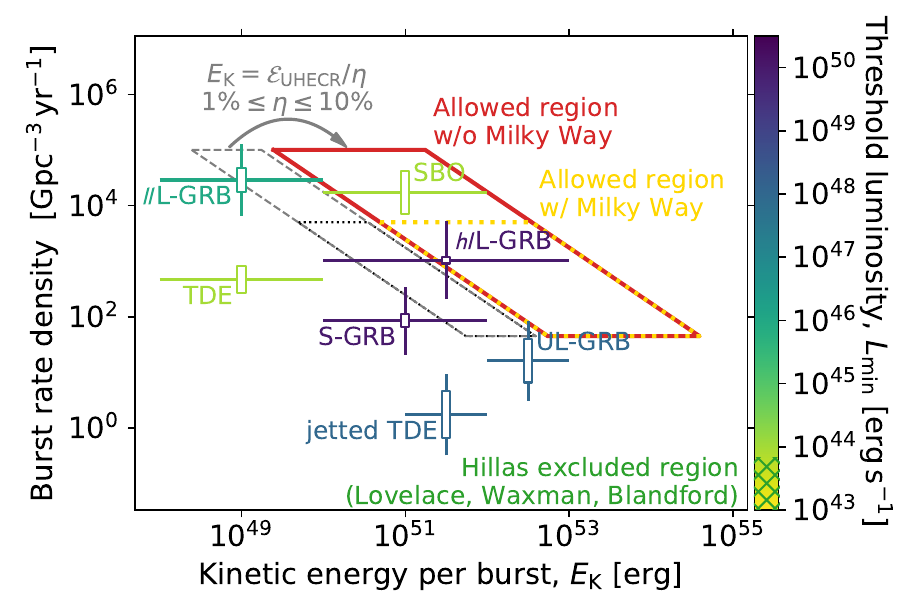}
\caption{UHECR burst rate as a function of the cosmic-ray energy emitted per burst (\textit{left}), and as a function of kinetic energy of the outflow (\textit{right}). The solid-red and dotted-gold lines delimit the regions allowed by the composition, spectrum  and arrival-direction data when including or not transient emission from the the Milky Way in the model. On the left-hand side, the colors of the two y-axes show the burst rate per unit matter obtained for UHECR sources evolving as SFRD or SMD. On the right-hand side, the allowed regions are based on an efficiency of conversion of kinetic energy to particles, $\eta$, between $1\,\%$ and $10\,\%$. Rectangular markers indicate the statistical uncertainty associated with the number of X-ray bursts observed for each type of sources. The vertical error bars show the range associated with the beaming correction factor, while the horizontal error bars illustrate the range of kinetic energy per burst in each population. The color bar shows the threshold luminosity above which each source density is measured. The low luminosities excluded by the Hillas-Lovelace-Waxman-Blandford criterion is shown as a green hatched region of the color bar.}
\label{fig:constraints}
\end{figure*}

The mass fraction of heavy nuclei (above Ne) to intermediate nuclei (He and CNO) is estimated as 
\begin{align}
    \label{eq:abundance_ratio1}
    \left.\frac{M(\mathrm{Ne}-\mathrm{Fe})}{M(\mathrm{He-O})}\right|_\mathrm{UHECR} 
    &= \frac{\mathcal{E}_\mathrm{Si}+\mathcal{E}_\mathrm{Fe}\frac{A_\mathrm{Fe}}{2Z_\mathrm{Fe}}}{\mathcal{E}_\mathrm{He}+\mathcal{E}_\mathrm{CNO}} \nonumber \\
    & =  0.30 \pm 0.05_\mathrm{stat.} \pm 0.10_\mathrm{sys.}.
\end{align}
A stellar-sized transient scenario should then account for a proportion of Ne$-$Fe elements (synthesized by C-, O- and Si-burning) to CNO elements (synthesized by the CNO cycle and He-burning) at least as large as the value in Equation~\eqref{eq:abundance_ratio1}. This is in line with the ratio of heavy to CNO nuclei dispersed by the explosions of massive stars in the interstellar medium, namely $\nicefrac{M(\mathrm{Ne}-\mathrm{Fe})}{M(\mathrm{CNO})} = 0.53 \pm 0.09$ (\citealt{2009LanB...4B..712L}, see also \citealt{2018PhRvD..97h3010Z} for more detailed accretion-ejection models).\footnote{The systematic uncertainty on the quantity estimated in Equation~\eqref{eq:abundance_ratio1} is smaller than that on the $\nicefrac{M(\mathrm{Ne}-\mathrm{Fe})}{M(\mathrm{CNO})}$ ratio, as can be seen by comparing the CNO proportions for the SMD and SFRD evolution scenarios.}

The elemental distribution at the sources  appears to be consistent with that of massive stars stripped of their outer H and He layers. On the other hand, the He-poor composition inferred in Equation~\eqref{eq:abundance_ratio2} rules out the acceleration of ionized atoms from the interstellar or intergalactic medium, as might be expected, e.g., in the termination shocks of jets from active galactic nuclei \citep[see, e.g.,][]{2019MNRAS.482.4303M}. The composition argument alone does not exclude the acceleration, within these jets, of ejecta from the disruption or explosion of a massive star \citep[see, e.g.,][for recent references]{2017PhRvD..96f3007Z, 2018NatSR...810828B,2023A&A...677L..14B}. Consideration of the rate and energetics of UHECR bursts provided in the next section, however, narrows down the populations of astrophysical sources that are candidates for ultra-high-energy acceleration, at least for those populations whose burst rate density is well constrained by observations.

\subsection{Burst rate and kinetic energy}

As shown in Figure~\ref{fig:scan_k}, arrival-direction data constrain the UHECR burst rate per unit SFR or $M_\star$. Such constraints are illustrated by the horizontal bands in Figure~\ref{fig:constraints}, left, which include systematic uncertainties in SFR and $M_\star$. The constraints are fairly independent of the tracer normalized relative to the SFR and $M_\star$ density measurements by \cite{2018MNRAS.475.2891D} up to a redshift $z=0.08$. A more direct comparison between the model and the UHECR arrival directions than that presented in Section~\ref{sec:median_maps}, e.g.\ using an approach similar to that employed by \cite{PierreAuger:2018qvk, PierreAuger:2022axr}, is expected not only to yield a correlated signal fraction larger than that inferred with the more limited catalogs tested thus far (all the galaxies in the volume are virtually considered in the approach presented here) but also to restrict even more tightly the parameter space. The latter is limited here by the mere existence of hotspots in the two prominent sky regions identified by the Pierre Auger and Telescope Array collaborations. 

In Figure~\ref{fig:constraints}, left, the diagonal bands show the constraints on the energy production rate, $\mathcal{L} \propto k \mathcal{E}_\mathrm{UHECR}$ with $\mathcal{E}_\mathrm{UHECR} = \sum_A \mathcal{E}_A$, as derived from the fit of spectral and composition data (see Section~\ref{sec:spectrum_compo_bestfit}). These diagonal bands take into account the systematic uncertainty in the energy scale, which is a dominant factor due to the rapid decrease of the UHECR flux with energy. Combining the constraints from spectral and composition data with those from arrival directions partially breaks the degeneracy between burst-rate density and the quantity of UHECRs emitted per burst. The resulting constraints are represented by the gold and red bands in Figure~\ref{fig:constraints}, left, which are obtained by including or not including transient emission from the Milky Way in the UHECR sky model, respectively (see Section~\ref{sec:median_maps}).

\begin{deluxetable*}{cccccccc}
\tablecaption{Properties of the stellar-sized X-ray transients. \label{tab:src_cand}}
\tablehead{
\colhead{Type} & \colhead{Beam angle} & \colhead{$\log_{10} E_\mathrm{K}$} & \colhead{Reference}	& \colhead{$\log_{10}  L_\mathrm{min}$} & \colhead{$\log_{10} \dot n_\mathrm{obs}$} & \colhead{Reference} & \colhead{$\log_{10} \dot n_\mathrm{true}$}\\
 & deg & [erg] &	& [erg$\,$s$^{-1}$] & [Gpc$^{-3}\,$yr$^{-1}$] &  & [Gpc$^{-3}\,$yr$^{-1}$]
}
\startdata
\textit{ll}L-GRBs	 & $5-20$ & $48	- 50$ & \cite{2017AdAst2017E...5C} & $46$ & $\phantom{-}2.64 \pm 0.21$ & \cite{2015ApJ...812...33S} & $4.5 \pm 0.6$ \\
SBOs	 & $-\phantom{0}$ & $50	-	52$ & \cite{2017hsn..book..967W} & $44$ & $\phantom{-}4.24 \pm 0.39$ & \cite{2015ApJ...812...33S} & $4.2 \pm 0.4$ \\
\textit{hl}L-GRBs	 & $1-5\phantom{0}$ & $50	-	53$ & \cite{2017AdAst2017E...5C} & $50$ & $-0.10 \pm 0.06$ & \cite{2015ApJ...812...33S} & $3.0 \pm 0.7$ \\
TDEs	 & $-\phantom{0}$ & $48	-	50$ & \cite{2022ApJ...938...28C} & $44$ & $\phantom{-}2.67 \pm 0.24$ & \cite{2015ApJ...812...33S} & $2.7 \pm 0.2$ \\
S-GRBs	 & $5-20$ & $50	-	52$ & \cite{2022ApJ...935L..11L} & $50$ & $\phantom{-}0.12 \pm 0.12$ & \cite{2015ApJ...812...33S} & $1.9 \pm 0.6$ \\
UL-GRBs	 & $5-20$ & $52	-	53$ & \cite{2015MNRAS.454.1073B} & $48$ & $-0.61 \pm 0.39$ & \cite{2017MNRAS.464.3568P} & $1.2 \pm 0.7$ \\
Jetted TDEs	 & $5-20$ & $51	-	52$ & \cite{2022ApJ...938...28C} & $48$ & $-1.58 \pm 0.42$ & \cite{2015ApJ...812...33S} & $0.2 \pm 0.7$
\enddata
\tablecomments{In column 8, the true rate density of bursts $\dot n_\mathrm{true} = f_\mathrm{b}^{-1} \dot n_\mathrm{obs}$ is determined from the rate observed in soft and hard X-rays, $\dot n_\mathrm{obs}$ (column 6), above a luminosity threshold $L_\mathrm{min}$ (column 5). The true rate density also accounts for the beaming correction factor of the relativistic component, $f_\mathrm{b} = 1 - \cos \theta_\mathrm{jet}$, where $\theta_\mathrm{jet}$ is the two-sided jet opening angle, or beam angle (column 2). The latter is taken as $90^\circ$ for non-collimated outflows from SBOs and TDEs.}
\end{deluxetable*}

Assuming a conversion efficiency of the kinetic energy or Poynting flux of the ejecta into cosmic rays,  $\eta = 1-10\,\%$, we bracket in Figure~\ref{fig:constraints}, right, the burst rate density and kinetic energy per burst to constrain the populations of candidate sources. We use the kinetic energy, $E_\mathrm{K}$, of the ejecta rather than its electromagnetic fluence, since the former is more directly related to the total cosmic-ray energy of each burst through $\mathcal{E}_\mathrm{UHECR} = \eta E_\mathrm{K}$.

Soft and hard X-ray observations, mainly from \textit{Swift}, have been instrumental in characterizing the properties of the populations of stellar-sized explosions and mergers listed in Table~\ref{tab:src_cand}, namely supernova shock breakouts (SBOs), non-jetted and jetted tidal disruption events (TDEs), short-duration gamma-ray bursts (S-GRBs) and long-duration gamma-ray bursts (L-GRBs). The latter are divided in Figure~\ref{fig:constraints} into sub-categories: high-luminosity and low-luminosity (\textit{ll}L-GRBs and \textit{hl}L-GRBs) for L-GRBs above and below $10^{50}\,$erg\,s$^{-1}$, as well as ultra-long GRBs, whose prompt emission lasts several ks (see references in Table~\ref{tab:src_cand}). As indicated by the color bar in Figure~\ref{fig:constraints}, right, the luminosity of non-jetted TDEs and SBOs hardly meets the Hillas-Lovelace-Waxman-Blandford criterion in Equation~\eqref{eqn:Hillas_Lovelace}. Jetted TDEs appear to be too infrequent to explain the UHECR sky. The kinetic energy of the ejecta of S-GRBs appears to be too low, although tighter constraints on their outflows may be needed to formally settle the matter. The various classes of L-GRBs seem to satisfy all the criteria presented in Figure~\ref{fig:constraints}, right. We note that \textit{hl}L-GRBs occupy a central position in the parameter space. 

\newpage

\section{Conclusion}

In this study, we develop a three-dimensional model of the UHECR sky aimed at reproducing the main features of the measurements from the Pierre Auger Observatory and Telescope Array in the flux suppression region, i.e.\ above 40\,EeV. In this energy range, excesses on Gaussian angular scales of the order of $15^\circ$ are emerging, as indicated by pieces of evidence between $4.2\,\sigma$ and $4.6\,\sigma$ provided by the Pierre Auger Collaboration  \citep{PierreAuger:2022axr} and the joint working group with the Telescope Array Collaboration \citep{PierreAuger:2023mvf}. The evidence comes from a $10-20\,\%$ anisotropic excess of events correlated with a flux-limited catalog of about 40 star-forming galaxies. The model proposed here takes advantage of the 400\,000-galaxy catalog of \cite{Biteau:2021pru} and of the tensor propagation formalism discussed in Section~\ref{sec:spectrum_compo_bestfit} that allow us to efficiently capture close to 100\,\% of the UHECR flux expected above 40\,EeV, irrespective of the cosmological evolution of the sources. The fit of a generic model to composition and spectrum data from ${\simeq}\,1$ to ${\simeq}\,100\,$EeV constrains the UHECR emissivity per unit matter as well as the ratio of heavy elements to intermediate-mass nuclei at escape from the sources. The latter suggests the ejecta composition of short-lived massive stellar progenitors, which supports a UHECR origin in transient stellar-sized sources. The software that enables us to efficiently fit spectral and composition data and to produce flux maps is made publicly available \citep{condorelli_2024_11440864}.

In addition to the three spatial dimensions covered with unprecedented depth at ultra-high energies, this study also accounts for the temporal dimension. We consistently consider the temporal and angular spreads caused by the magnetic fields that UHECRs cross. We identify the parameter space region where the median transient sky model shows an excess in the Centaurus region and an excess in the Ursa Major region, as suggested by UHECR observations above 40\,EeV. The flux contrast and angular spread inferred from the observations are found in this region of parameter space.

The nodal point of the model, which makes it falsifiable, is the existence of a turbulent magnetic field in the Local Sheet. Its coherence length of 10\,kpc and amplitude of $0.5-20$\,nG match the properties inferred for galaxy filaments \citep{2021MNRAS.505.4178V,2022MNRAS.512..945C}. Whether radio observatories such as LOFAR could already probe the signatures of such a field, or whether future measurements with the Square Kilometer Array could reveal it, remains to be determined \citep[e.g.][for a review]{2018PASJ...70R...2A}. Our elementary model deliberately leaves out the coherent fields, the specific realization of which necessarily affects the comparison between model and UHECR data \citep[e.g.][for recent discussions]{2018ApJ...861....3B, 2018MNRAS.475.2519H, 2021ApJ...913L..13D, 2023arXiv231112120U}. We discuss an example of factoring in coherent Galactic deflection models in Appendix~\ref{sec:gmf}.

The presence of excesses in the Centaurus and Ursa Major regions allows us to bracket the burst rate per unit matter. This, combined with the production rate of UHECRs, constrains the total energy emitted by each burst of UHECRs and hence the kinetic energy of the outflows within which acceleration takes place. The conclusion is that, among the stellar-sized transient X-ray sources with a measured rate density, only the L-GRB population can be the source of all UHECRs. The discovery of the sources requires the confirmation of the anisotropy of the UHECR sky above 40\,EeV with $5\,\sigma$ confidence level.

More precise measurements of the properties of populations of non-thermal transient sources will certainly yield a sharper picture of the L-GRB subtype favored for UHECR acceleration. Unprecedented exploration of transient populations is expected with the development of time-domain astronomy, both at the highest observed photon energies with the Cherenkov Telescope Array Observatory, in X-rays with satellites such as Einstein Probe, and at optical wavelengths with the Vera C.\ Rubin Observatory. These observations will also shed light on the acceleration, radiation and escape processes that take place in transient astrophysical sources \citep[e.g.][and references therein]{2019ARNPS..69..477M}, subjects we have deliberately left aside in this macroscopic study of the cosmic-ray sky at the highest energies.

We have limited the scope of this work to modeling the cosmic-ray sky in the flux-suppression region beyond 40\,EeV, where our galaxy catalog captures over 97\,\% of the expected flux. The lower-energy region, down to the ankle of the spectrum at 5\,EeV, also presents particularly interesting observational features \citep{PierreAuger:2017pzq, PierreAuger:2020fbi}. The observation of a flux dipole of increasing amplitude from $4-8$\,EeV to $16-32$\,EeV has been the subject of numerous models, all suggesting qualitative agreement with a distribution of sources following that of large-scale structures in the local universe \citep{2015PhRvD..92f3014H, 2018ApJ...854L...3W, 2018MNRAS.475.2519H, 2022A&A...664A.120A, 2024ApJ...966...71B}. Our model, like the others mentioned above, yields an increase in dipole amplitude with increasing energy. However, measurements suggest that cosmic-ray rigidity is two-to-three times lower at 8\,EeV than in the flux-suppressed region \citep{2019FrASS...6...23B}, so that angular deviations are expected to be two-to-three times larger. The coherent components of the magnetic fields traversed therefore appear to be an essential ingredient for fully modeling the UHECR dipole. In addition to the advances on this front from radio astronomy, the SPHEREx infrared satellite has the potential to provide a complete spectroscopic mapping of the nearby universe, and thus to deepen the reconstruction of the source distribution responsible for the UHECR dipole. 

The constraints drawn on extragalactic UHECR accelerators may also have consequences for cosmic-ray nuclei observed below the ankle energy. With rare transient events in each galaxy, we have seen that the non-observation of trans-EeV particles from past acceleration episodes in the Milky Way further shrinks the allowed range of burst rate per unit matter. 
As the SFR of the Milky Way is estimated at $\mathrm{SFR}_{\mathrm{MW}}\simeq 1.6\,M_\odot\,\mathrm{yr}^{-1}$~\citep{2015ApJ...806...96L}, our constraints result in a Galactic burst rate of $k\mathrm{SFR}_\mathrm{MW}\simeq 1 \cdot 10^{-5} - 3 \cdot 10^{-3}$\,yr$^{-1}$, which brackets the estimate reported in~\cite{2013arXiv1311.0287K,2002astro.ph..5272L}. Further investigations of the Galactic confinement of cosmic rays are needed to explore the viability of rare Galactic events to fuel the energy density of sub-ankle nuclei; they are left for future studies.

\acknowledgments 

We acknowledge the work of the last two reviewers. Their constructive comments helped to improve the form and content of this manuscript. We are also very grateful to the colleagues who kindly provided comments on this manuscript, in particular Martin Lemoine, Kohta Murase, Peter Tyniakov and Roger Clay, as a member of the Pierre Auger Collaboration publication committee.
We gratefully acknowledge funding from ANR via the grant MultI-messenger probe of Cosmic Ray Origins (MICRO), ANR-20-CE92-0052.  This work was also made possible with the support of the Institut Pascal at Universit\'e Paris-Saclay during the Paris-Saclay Astroparticle Symposium 2022, with the support of the P2IO Laboratory of Excellence (program ``Investissements d’avenir'' ANR-11-IDEX-0003-01 Paris-Saclay and ANR-10-LABX-0038), the P2I axis of the Graduate School Physics of Universit\'e Paris-Saclay, as well as IJCLab, CEA, IPhT, APPEC, the IN2P3 master projet UCMN, EuCAPT ANR-11-IDEX-0003-01 and Paris-Saclay ANR-10-LABX-0038). 

\newpage
\appendix

\section{Impact of Galactic Magnetic Field} 
\label{sec:gmf}
 
When constructing the sky maps, the impact of the magnetic fields has been modeled by spreading the flux from a source over an angular scale $\Theta$. This is an approximation that neglects mass-spectrometric-type as well as flux-magnification effects, which can become important particularly in the case of large-scale fields such as the Galactic one. We show here the extent to which this approximation proves relevant.

We follow \cite{Jansson2012AField} for the Galactic magnetic field and consider a Kolmogorov turbulence with an intensity scaled to the coherent component, and a turbulence alone for the Local Sheet. The deflections from the turbulence follow from \cite{1998tx19.confE.617A}. To build the sky maps, we perform numerical experiments using $10^6$ test particles. Cosmic-ray particles with opposite charge are back-tracked from their arrival direction $\nn_k$ observed on Earth, drawn at random from a uniform distribution on the sphere. Arrival directions $\nn_k$ are then given a weight proportional to those of the sources sampled within $0.5^\circ$ along the trajectory of the test particles. Using a back-tracking approach, the directional-dependent effects of flux magnification, defined as the flux received from a given source relative to that received in the absence of the magnetic field, are handled automatically. In this respect, our results are consistent with those of \cite{Farrar:2014hma} and \cite{Farrar:2017lhm}. This would not be the case should we forward-track particles from sources to Earth by means of ``lookup tables'' to map the entry point in the Galaxy with the arrival direction of a single UHECR on Earth, unless magnification factors are introduced into the lookup tables as in \cite{Bretz:2013oka}. We note that the magnification effects stemming from the turbulence may be overlooked here. The geometry of the turbulence in the Galaxy may in addition not be accurately described by an homogeneous and isotropic power spectrum, as an anisotropic one \citep{Goldreich:1994zz} has been shown to describe better, at least locally, the cosmic-ray anisotropies in the TeV--PeV energy range~\citep{Giacinti:2016tld}. A full account of these effects requires to track test-particles in specific realizations of turbulence, beyond the application of random deflections. Such a comprehensive study is beyond the scope of the present work and will be reported elsewhere.

\begin{figure}[t!]
\centering
\includegraphics[width=0.7\columnwidth]{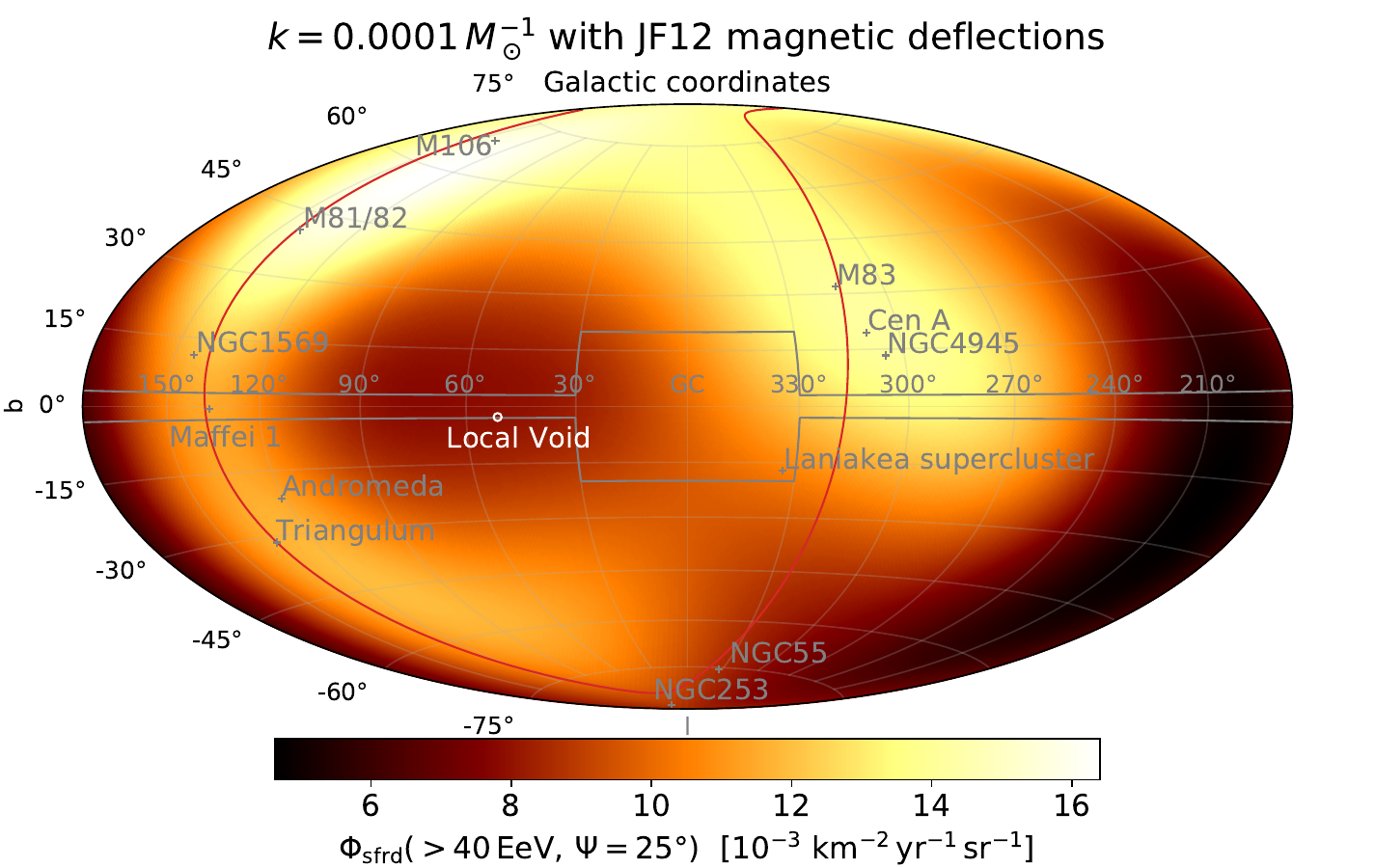}
\caption{Model flux map of UHECRs at energies above 40\,EeV including deflections in the model of \cite{Jansson2012AField} of the Galactic magnetic field, for a burst rate per unit SFR $k = 1\times10^{-4}\,M_\odot^{-1}$. The sky map is smoothed with a top-hat function of radius $\Psi = 25\,^\circ$ to ensure a meaningful comparison with the observations. The red line shows the supergalactic plane. The area delimited by the grey line represents the zone of avoidance.}
\label{fig:gmfimpact}
\end{figure}

The exact content in each element received on Earth impacts the results due to the rigidity dependences of both the coherent deflections and of the flux-magnification effects in a large-scale structured magnetic field. We consider for simplicity the benchmark mass composition as inferred from the best-fit parameters. We show the results of this procedure in Figure~\ref{fig:gmfimpact}. A comparison with Figure~\ref{fig:modelmaps} shows that the main changes induced by the large-scale component of the Galactic magnetic field are a flux magnification of the UMa region (around M\,81/82) and a pattern distortion along the supergalactic plane from NGC\,1569 to NGC\,253. With caution regarding both the model of the Galactic magnetic field and the flux determination currently obtained in the UMa region with the Telescope Array exposure, we note that flux magnification from the model results in a flux contrast between the UMa and Cen regions that may better match observations. On the other hand, for a burst rate $k = 1\times10^{-4}\,M_\odot^{-1}$, the model suggests a flux excess between NGC\,1569 and NGC\,253 at the level of ${\simeq}\, 12\times10^{-3}~$km$^{-2}~$yr$^{-1}~$sr$^{-1}$, which remains difficult to uncover with the current exposures of the Pierre Auger Observatory and Telescope Array in this part of the sky.

The exploratory study case discussed here shows the expected impact of the large-scale component of the Galactic magnetic field, the knowledge of which still suffers from large uncertainties. It is observed that both mass-spectrometric-type and flux-magnification effects yield a sky map that is consistent with the current observational data.

\newpage
\
\vfill
\section{Skymaps in the figure set (arXiv version)} 
\label{sec:arXiv}

\begin{figure*}[ht!]
\includegraphics[width=0.49\textwidth]{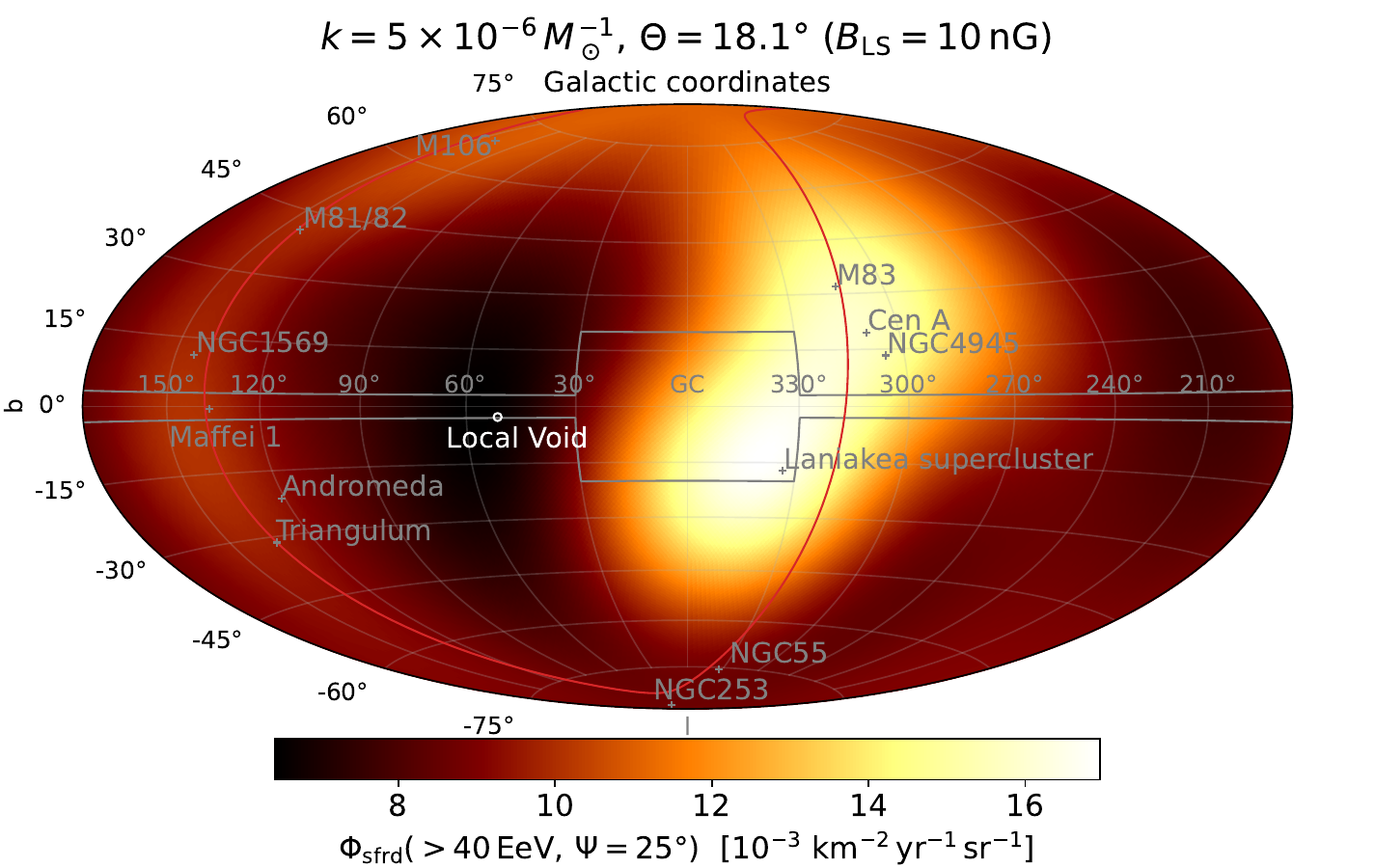}
\includegraphics[width=0.49\textwidth]{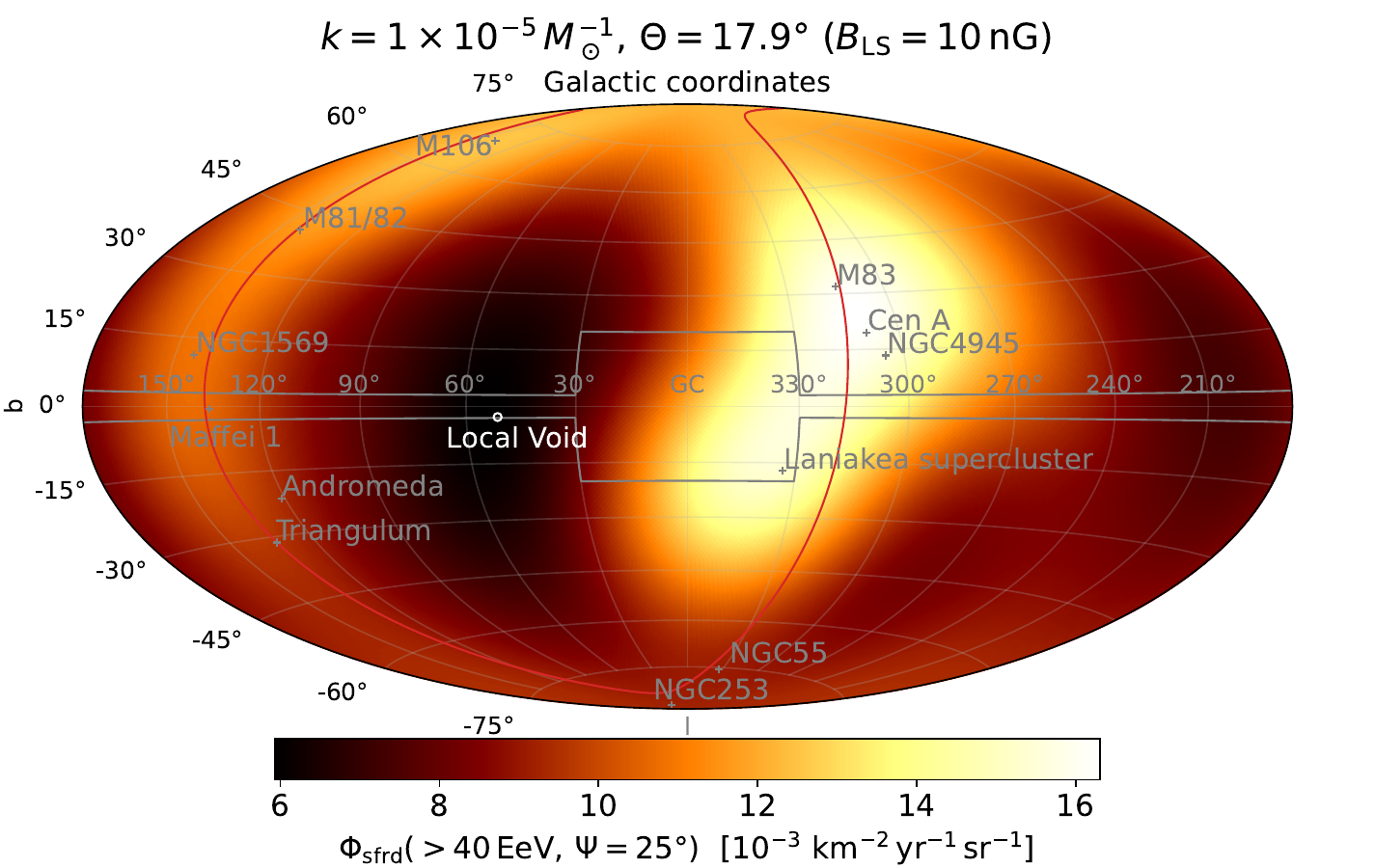}
\includegraphics[width=0.49\textwidth]{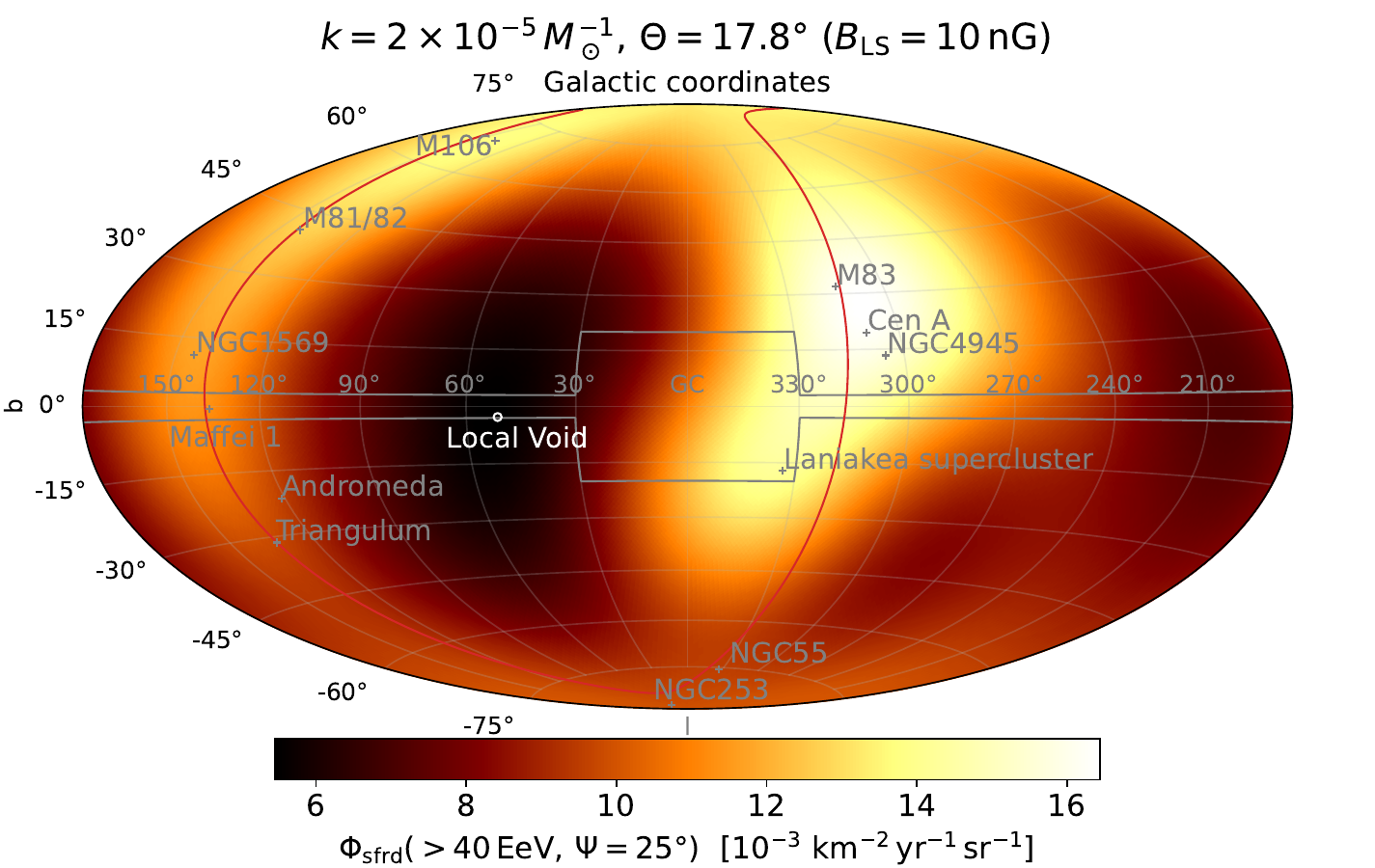}
\includegraphics[width=0.49\textwidth]{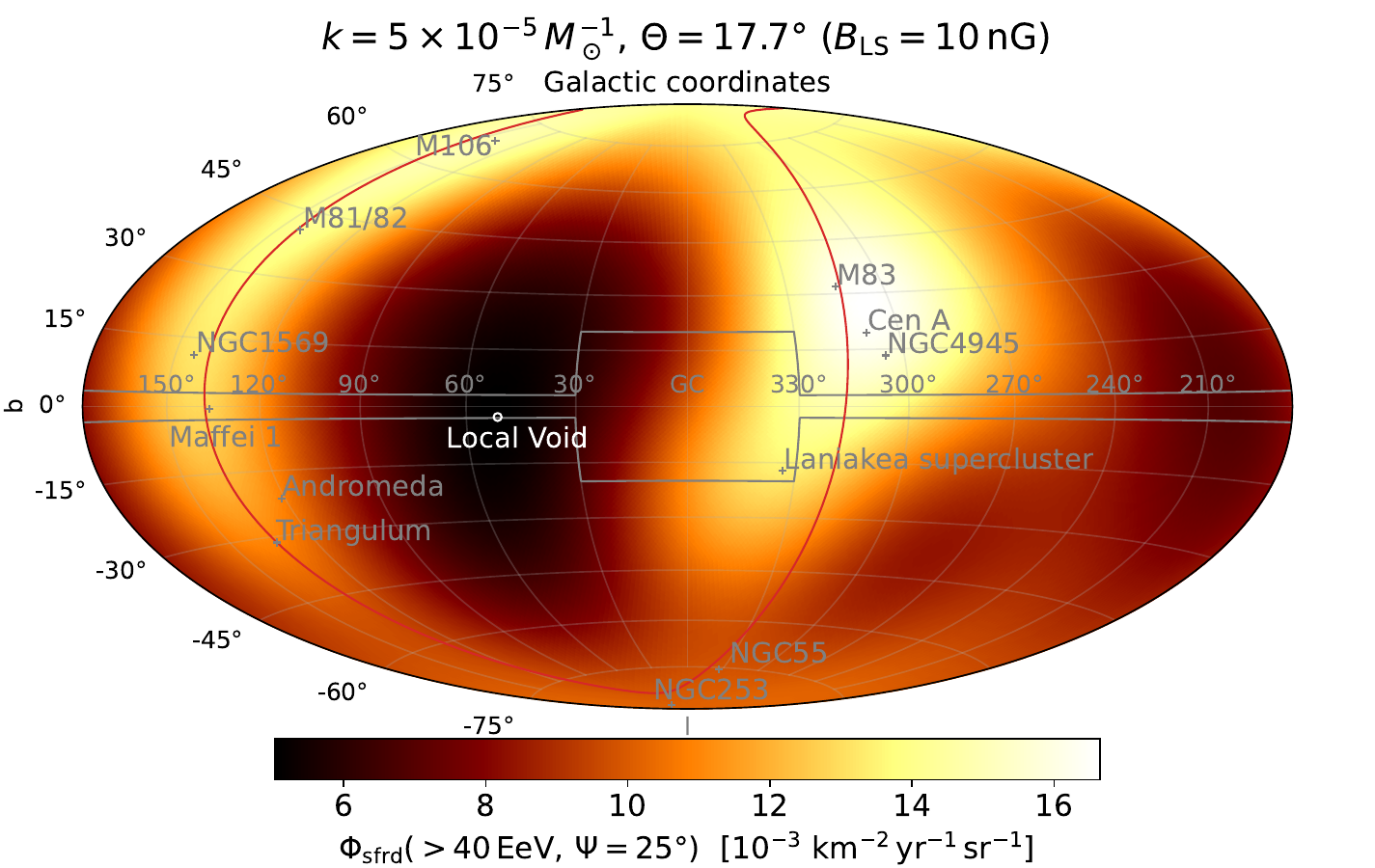}
\includegraphics[width=0.49\textwidth]{uhecr_fluxmap_19.6_sfrd_10nG_e.pdf}
\includegraphics[width=0.49\textwidth]{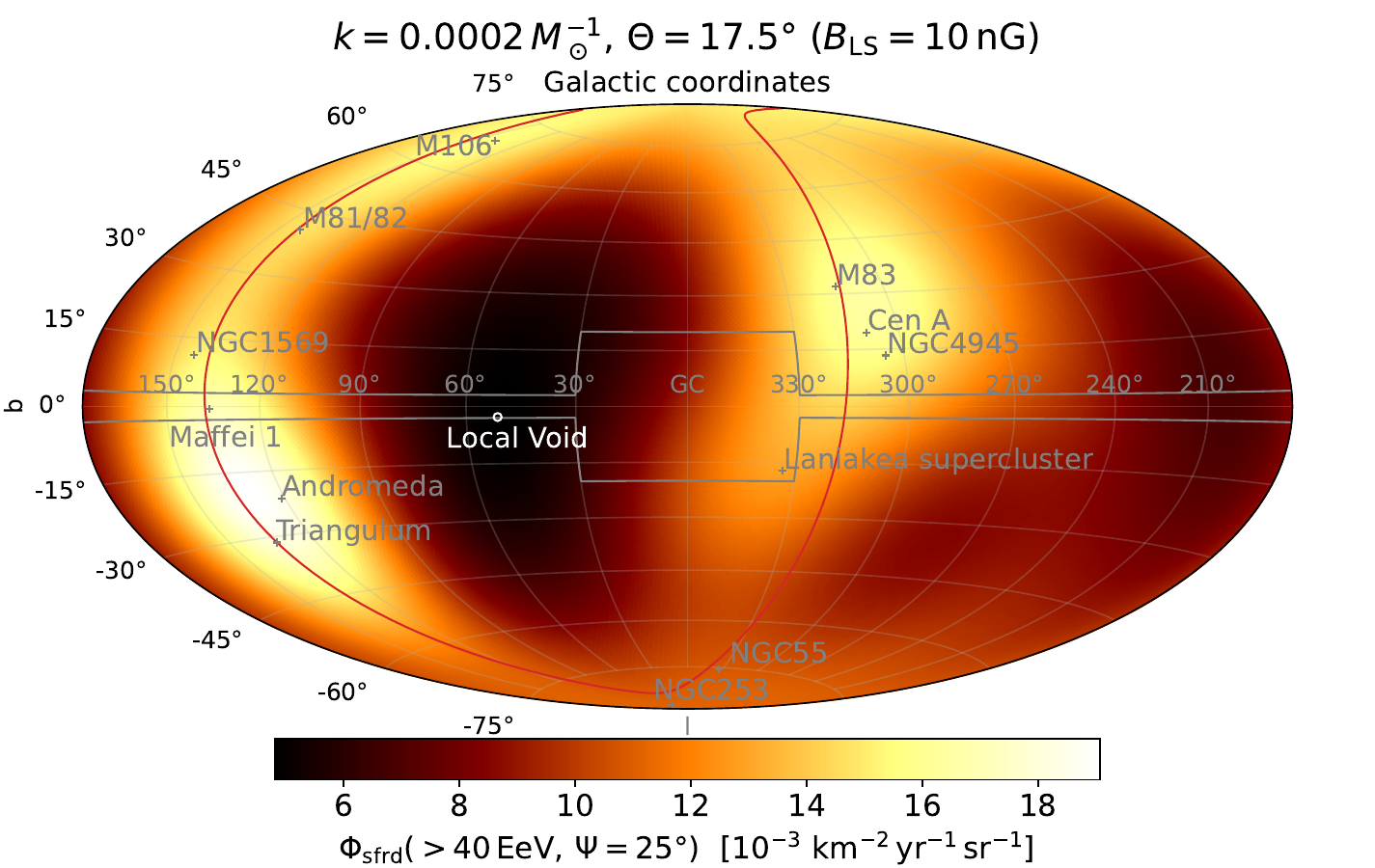}
\includegraphics[width=0.49\textwidth]{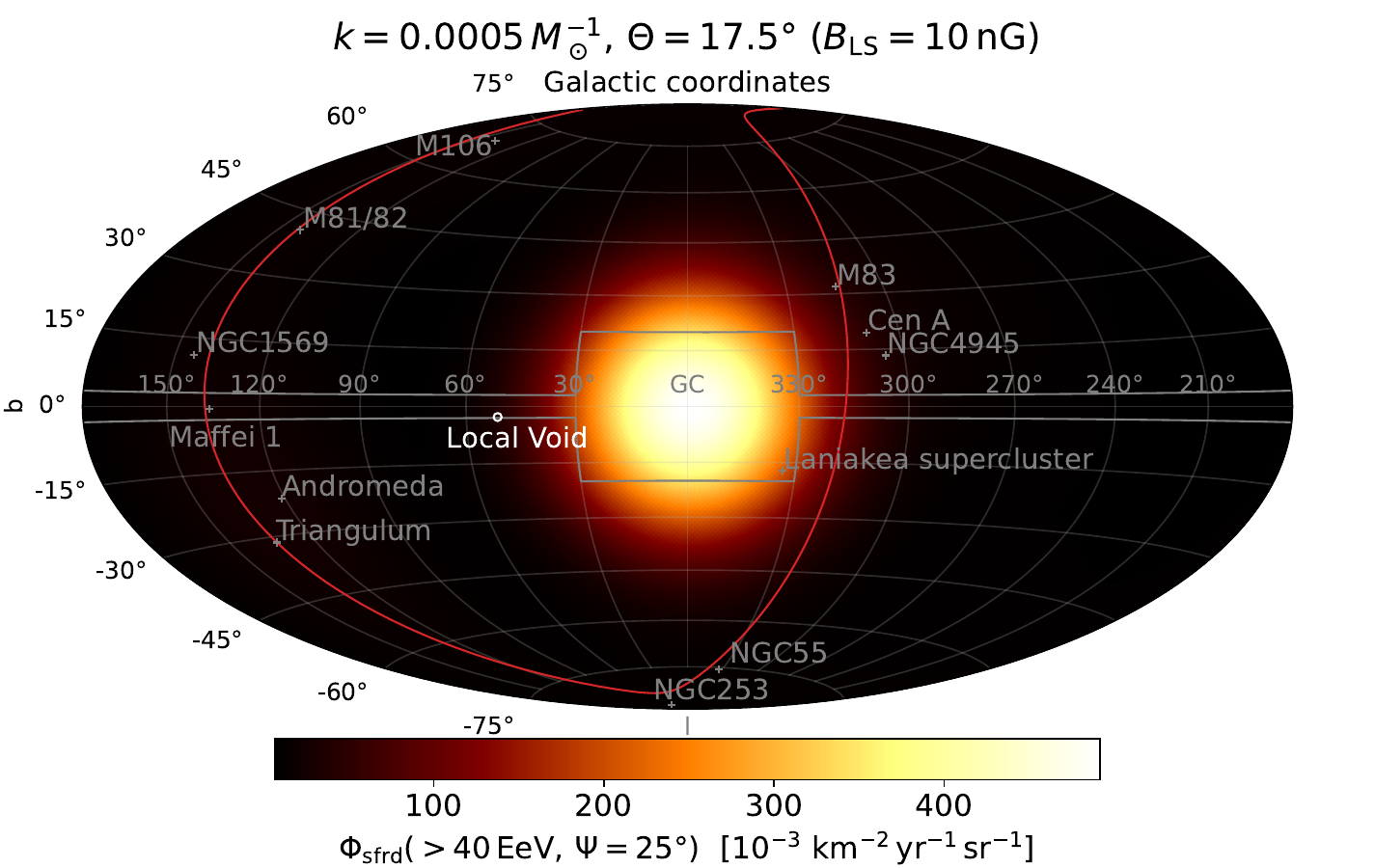}
\caption{Model flux map of UHECRs at energies above 40\,EeV for a burst rate per unit SFR $k$ ranging from $5\times10^{-6}\,M_\odot^{-1}$ to $5\times10^{-4}\,M_\odot^{-1}$. The sky map are smoothed using a top-hat function of radius $\Psi = 25\,^\circ$ for comparison with observational data.}
\label{fig:modelmaps_app}
\end{figure*}

As the set of figures~\ref{fig:modelmaps} is only available in the online version of the journal, we show in Figure~\ref{fig:modelmaps_app} the model flux maps at energies above 40\,EeV for a burst rate per unit SFR $k$ ranging from $5\times10^{-6}\,M_\odot^{-1}$ to $5\times10^{-4}\,M_\odot^{-1}$, with boundaries of this range excluded by the criterion discussed in Section~\ref{sec:median_maps}. At the lowest $k$ value, our supercluster Laniakea dominates the sky. Larger values of $k$ reveal galaxies in the Local Sheet, and even in the Local Group for the highest allowed values (see map for $k = 2\times10^{-4}\,M_\odot^{-1}$). The emission from the Milky Way is included here in the model as a transient point source colocated with Sgr A*. Our Galaxy dominates the sky for $k \geq 5\times10^{-4}\,M_\odot^{-1}$.

\newpage

\bibliographystyle{aasjournal.bst}
\bibliography{bibliography}

\end{document}